\documentclass[11pt]{article}
\pdfoutput=1

\usepackage[]{Packages}
\bibliography{T-duality}

\usepackage{Definitions}

\usepackage{textcomp}

\preprint{\texttt{IPhT-t15/nnn}\\
  \texttt{IPMU15-0040}}

\newcommand{\OfficialTitle}{4D/3D reduction of dualities: mirrors on the circle}

\title{\vspace{2cm}
  {\Huge\textbf{\dosserif\OfficialTitle}}
}

\hypersetup{pdfauthor={Antonio Amariti and Davide Forcella and Claudius Klare and Domenico Orlando and Susanne Reffert},pdftitle={\OfficialTitle}}

\author{%
  \begin{minipage}{.8\linewidth}
    \vspace{1cm}
    \begin{center} \dosserif
      {\small 
        \textbf{Antonio~Amariti}\textsuperscript{$\spadesuit$},
        \textbf{Davide~Forcella}\textsuperscript{$\spadesuit\spadesuit$}, 
        \textbf{Claudius~Klare}\textsuperscript{$\heartsuit$}, 
        \textbf{Domenico~Orlando}\textsuperscript{$\spadesuit,\clubsuit$} and 
        \textbf{Susanne~Reffert}\textsuperscript{$\diamondsuit$}}
    \end{center}
    \vspace{1cm}
    \authorBlock{$\spadesuit$}{\textsc{lptens} -- \textsc{umr cnrs} 8549, 24, rue Lhomond, 75231 Paris, France} 
    \authorBlock{$\spadesuit\spadesuit$}{Physique Th\'eorique et Math\'ematique and International Solvay Institutes\\ \textsc{ulb}, C.P. 231, 1050 Bruxelles, Belgium}
    \authorBlock{$\heartsuit$}{\textsc{ipht}, \textsc{cea}/Saclay, 91191 Gif-sur-Yvette, France}
    \authorBlock{$\heartsuit$}{\textsc{ihes}, 35, Route de Chartres, 91440 Bures-sur-Yvette, France}
    \authorBlock{$\clubsuit$}{\textsc{ipt} Ph. Meyer, 24, rue Lhomond, 75231 Paris, France}
    \authorBlock{$\diamondsuit$}{\textsc{itp -- aec}, University of Bern, Sidlerstrasse 5, 3012 Bern, Switzerland}
  \end{minipage}
}

\date{}

\begin{document}

\setstretch{1.15}

\numberwithin{equation}{section}

\begin{titlepage}

  \newgeometry{top=23.1mm,bottom=46.1mm,left=34.6mm,right=34.6mm}

  \maketitle

  \thispagestyle{empty}

  \vfill\dosserif
  
  \abstract{\normalfont \noindent
We engineer a brane picture for the reduction of Seiberg dualities
from $4$D to $3$D, valid also in the presence of orientifold planes.
We obtain effective $3$D dualities on the circle
by T--duality, geometrizing the non-perturbative superpotential which is an affine Toda potential.
When reducing to pure $3$D, we define a double-scaling limit 
which creates a sector of interacting singlets, giving a unified mechanism for the brane reduction of dualities.
  }

  \vfill

\end{titlepage}

\restoregeometry




\section{Introduction}

We construct the brane representation for the reduction of gauge theory dualities from $4$D to $3$D. 
This analysis was started in~\cite{Amariti:2015yea}
where it was shown how to translate the dimensional  reduction of dualities in
terms of \D{}-- and \NS{}--branes in \tIIA supergravity.
Here we elaborate on this picture, finding an algebraic description of the superpotential involved in the dimensional reduction
which incorporates the generalization from unitary to real gauge groups.
The brane picture gives a unified treatment of various dualities involving different gauge groups and matter content.
It allows also to reduce further to pure $3$D dualities, by a double-scaling on the relative positions of some \D{}--branes and the radius of the circle.
In this process an extra sector is created in the magnetic theories,
reproducing the gauge theory duality in the pure $3$D limit in the brane picture.

\bigskip

Recent insight in the structure of supersymmetric field theories has been obtained relating results in different dimensions. 
One example is the similarity between the electric-magnetic duality discovered by Seiberg in~\cite{Seiberg:1994pq}
for four-dimensional \textsc{sqcd} and the three-dimensional dualities studied in~\cite{Karch:1997ux,Aharony:1997gp}.
The dualities can indeed be connected by dimensional reduction, as discussed in~\cite{Aharony:2013dha} (see also~\cite{Niarchos:2012ah}).
 It turns out that in a necessary intermediate step of this reduction one needs to consider the duality on $\mathbb{R}^3 \times S^1$.
 At scales lower than the inverse radius of $S^1$ this gives rise to a new, effective, $3$D duality. 
The presence of the circle is manifest through the contributions of \ac{kk} monopoles.
The limit to pure $3$D dualities, recovering \emph{e.g.\ }the results of~\cite{Karch:1997ux,Aharony:1997gp}, 
depends on the details of the gauge and matter content ~\cite{Aharony:2013dha,Aharony:2013kma}.

\bigskip 

\begin{figure}
  \centering
  \includegraphics{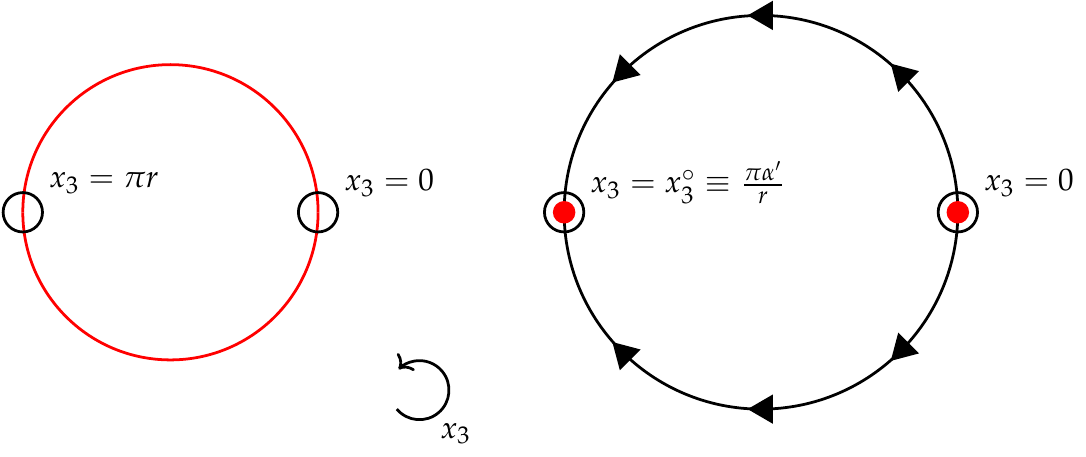}
  \caption{Geometry of the compact direction. The possible orientifolds are depicted in red color.
  LHS: \tIIA circle of radius $r$. 
  RHS: T--dual circle of radius $\alpha'/r$. The black arrowheads indicate the motion of the \D{}--branes to the mirror point 
  \(x_3 = x_3^\circ \equiv \pi \alpha'/r \).}
  \label{fig:new-orientifold}
\end{figure}

In this paper we study the brane construction of this reduction,
giving a physical origin for the differences in the pure $3$D limit.
We obtain the effective $3$D dualities by T--duality,
where Euclidean \D1--branes reproduce the non-perturbative effects of the \ac{kk} monopoles.
Equivalently these effects are captured by an algebraic formulation in terms of S--dual \F1--strings.
Configurations creating the monopole superpotential are classified by affine Dynkin diagrams and
this superpotential is an affine Toda potential.
As depicted in Figure~\ref{fig:new-orientifold}, we take the pure $3$D limit in the electric theory by moving some flavor branes to 
the mirror point \(x_3^\circ \), when sending the radius to infinity.
The magnetic dual is obtained by an \ac{hw} transition,
generating an additional gauge theory at $x^\circ_3$. 
This gauge theory is described by a sector of interacting singlets.
It is necessary in reproducing the limit to pure $3$D dualities.

The brane construction is quite general and can also be applied to theories with real gauge groups and tensor matter,
which require some extra treatment in the field theory analysis~\cite{Aharony:2013dha,Aharony:2013kma}.
In brane language these theories are obtained including orientifold planes.
Orientifolds are straightforwardly incorporated in our picture. 
When the theory is put on the circle, a second orientifold plane is generated after T-duality 
at the mirror point $x^\circ_3$~\cite{Hanany:2000fq,Hanany:2001iy}.
In the \acs{hw} transition the orientifold, carrying \D-brane charge, modifies the rank of the gauge groups both at $x_3=0$ 
and at $x_3=x^\circ_3$. By considering various brane realizations with orientifolds
we recover the $3$D dualities with orthogonal or symplectic gauge groups and those with tensor 
matter~\cite{Aharony:1997gp,Karch:1997ux,Giveon:2008zn,Niarchos:2008jb,
Orlando:2010uu,Orlando:2010aj,Willett:2011gp,Benini:2011mf,Kapustin:2011vz,Kim:2013cma,Aharony:2014uya}.

The plan of this article is as follows. 
In Section~\ref{sec:review},
we recap the reduction of $\mathcal{N}=1$ dualities from $4$ to $3$ dimensions.
In Section~\ref{sec:general}, we summarize the brane realization of the dimensional reduction.
We discuss the generation of an affine Toda potential on the Coulomb
branch variables when the theories are studied on a circle and its relation to the second orientifold appearing in the T--dual picture. 
Moreover, we explain the double-scaling limit and the reduction to pure three-dimensional dualities.
In Section~\ref{sec:symplectic} we apply the brane picture on symplectic gauge theories with fundamental, antisymmetric and adjoint matter.
In Section~\ref{sec:unitary-antisymmetric} we study the brane setup of unitary gauge groups with antisymmetric flavor.
In Section~\ref{sec:orthogonal} we study orthogonal gauge groups,
our analysis is at the level of the local properties and the gauge algebra.
We conclude in Section~\ref{ref:conclusions} by outlining some open questions.


\section{Brane reduction of dualities}

\subsection{Remarks on the 4D/3D reduction}
\label{sec:review}

In this subsection we review some general
field theoretical aspects of the reduction of \(\mathcal{N} = 1\) $4$D dualities to $\mathcal{N}=2$ dualities in $3$D.
For a more complete review see Section~2.1 of~\cite{Amariti:2015yea} and the original work~\cite{Aharony:2013dha}.
Here we just recall few aspects which are important for our analysis.

Connecting via dimensional reduction pairs of dual theories in $4$D to corresponding pairs in $3$D requires some care.
A consistent reduction has been obtained by studying the $4$D theory on a circle of finite radius $r$,
where a non-perturbative superpotential is generated from \ac{kk} monopoles on $S^1$.
We refer to this superpotential as $W_\eta$ in the rest of this paper.

In order to preserve the $4$D duality in the dimensional reduction where $r\rightarrow 0$,
one needs to consider, in some cases, an RG flow triggered by real masses of order $\frac{1}{r}$.
Furthermore, it has been argued in~\cite{Aharony:2013dha} that while some electric quarks are integrated out,
one sometimes has to consider the magnetic theory in a particular vacuum.
This magnetic vacuum corresponds to a large vev for the scalar field in the vector multiplet.

The details of the reduction -- involving the superpotential $W_\eta$,
the real mass flow and the non-trivial vacua -- depend on the nature of the gauge group and the matter content.
We refer the reader to the papers~\cite{Aharony:2013dha,Aharony:2013kma,Csaki:2014cwa,Nii:2014jsa,Amariti:2013qea} for more details and explicit examples. 

Let us note that the $4$D dualities on a finite circle,
with non-perturbative superpotentials $W_\eta$,
give rise to new, effectively 3 dimensional dualities, which are interesting in their own right.

\bigskip

To sum up, the aspects important for the forthcoming analysis are the superpotential $W_\eta$,
the real mass flow and the associated vacuum structure in the magnetic theory upon shrinking the circle to zero size.

In the following we describe the brane construction of this reduction.
We provide an algebraic description of $W_\eta$ in terms of the gauge group structure.
This allows for a generalization from unitary to real gauge groups.
We also find the brane description of the real mass flow and the vacuum structure.

\begin{table}
  \centering
  \begin{tabular}{lcccccccccc}
    \toprule
                & 0        & 1        & 2        & 3        & 4        & 5        & 6        & 7        & 8        & 9        \\ \midrule
    \D4         & $\times$ & $\times$ & $\times$ & $\times$ &          &          & $\times$ &          &          &          \\
    \D6         & $\times$ & $\times$ & $\times$ & $\times$ &          &          &          & $\times$ & $\times$ & $\times$ \\
    \D6'        & $\times$ & $\times$ & $\times$ & $\times$ & $\times$ & $\times$ &          & $\times$                       \\
    \NS{}       & $\times$ & $\times$ & $\times$ & $\times$ & $\times$ & $\times$ &          &          &          &          \\
    \NS'        & $\times$ & $\times$ & $\times$ & $\times$ &          &          &          &          & $\times$ & $\times$ \\
    \(\O4^\pm\) & $\times$ & $\times$ & $\times$ & $\times$ &          &          & $\times$                                  \\
    \(\O6^\pm\) & $\times$ & $\times$ & $\times$ & $\times$ & $\times$ & $\times$ &          & $\times$                       \\ \bottomrule
  \end{tabular}
  \caption{Brane setup for the realization of the gauge theories of interest in this paper.}
  \label{tab:O4+_branes}
\end{table}

\subsection{The general strategy}
\label{sec:general}

In this section we first summarize the brane engineering of theories with four supercharges on $\mathbb{R}^3 \times S^1$.
In the second part we provide a brane picture for reducing $4$D dualities to $3$D.
There, for the sake of being explicit, we focus on the example of $U(N)$ gauge theories and fundamental matter.
The analysis of real groups and tensor matter follows analogously
and is the subject of sections \ref{sec:symplectic}, \ref{sec:unitary-antisymmetric} and \ref{sec:orthogonal}.

A brane description of theories on $\mathbb{R}^3 \times S^1$ is found \emph{e.g.\ }in~\cite{Amariti:2015yea}.
Let us give a brief summary.
The four dimensional gauge theory is engineered by a \tIIA brane system of \D4--branes stretched between \NS{}-- and \NS'--branes
as denoted in Table \ref{tab:O4+_branes}.
A dimensional reduction  of field theories can be reproduced in this picture by T-dualizing along one compact,
space-like dimension (say $x_3$).
There is a compact Coulomb branch (\textsc{cb}) which is parameterized by the scalars $\sigma_i$ in the vector multiplet the dual photons $\phi_i$,
where $i=1\cdots \text{rank}(G)$.
The vev of the scalars $\sigma_i$, and hence the position on the \textsc{cb}, corresponds to the positions of the \D3-branes in $x_3$.
In this configuration the \D3 branes repel each other.
The force can be understood in terms of Euclidean \D1--branes stretched between the \NS~and the \D3--branes.

These \D1–branes map in the field theory to monopoles, which generate an \ac{ahw} superpotential for the \textsc{cb} coordinates.
The force of this superpotential maps to the repulsive force between branes.

Note that there is a special Euclidean \D1-brane when the configuration is compact, as depicted in the left side of Figure \ref{fig:Dynkin-An} 
(there the S-dual scenario with \F1-strings instead of \D1-branes is shown).
This special \D1-brane connects the $1$st and the $N$th \D3-brane ``on the rear side'' of the compact $x_3$.
In field theory this corresponds to an additional term in the superpotential.  
As mentioned in Section~\ref{sec:review}, a crucial role in the dimensional reduction of dualities is played by
the superpotential $W_\eta$, appearing at finite circle radius.
In the brane picture it is reproduced precisely by this special, winding \D1 brane.

In the literature the ``regular'' \D1-branes are usually referred to as \ac{bps}- and the ``special'' ones as \ac{kk}-monopoles.

\bigskip

So far the summary of engineering gauge dynamics from branes as given in~\cite{Amariti:2015yea}. 
Now we want to discuss how the superpotential $W_\eta$ is given in terms of gauge group data. 
Stable configurations of branes, possibly in the presence of orientifolds, 
are in one to one correspondence with the Dynkin diagram of the  gauge group. 
The Dynkin diagrams, in turn, are in one to one correspondence with the possible superpotentials $W_\eta$.

\begin{itemize}
  \item The fundamental (in the sense of~\cite{Weinberg:1979zt,Weinberg:1982ev}) \acs{bps} monopoles are labeled by the simple co-roots of the Lie co-algebra.
For unitary gauge groups $G$ this corresponds to placing the \(i\)--th \D1--brane between the \(i\)--th and the \((i+1)\)--th  \D3--brane (for $i=1,\dots,\rank(G)$).
It is useful to study the S--dual configurations where \D1--branes become \F1--strings.
In this picture the \D3--branes are still distributed on the circle and connected by \F1--strings, as depicted in the upper left corner of Figure~\ref{fig:Dynkin-An}.

We now exploit the crucial fact that
the spectrum of the allowed \acs{bps} \F1--strings is given by the simple roots of the corresponding Lie algebra~\cite{Garland:1988bv,Hanany:2001iy}.
For a unitary gauge group the simple roots of the $A_N$ series correspond to $\sigma_i-\sigma_{i+1}$,
\emph{i.e.} to the difference between the positions of two consecutive \D3--branes.

We can include real gauge groups by adding orientifold planes (see Appendix \ref{sec:orientifolds} for details).
The allowed spectrum of \F1--strings is again given by the corresponding Dynkin diagrams, classified by the \(B_N, C_N\) and \(D_N\) series.

  Summing the contributions from the \acs{bps} monopoles we obtain the superpotential on the Coulomb branch,
  finding a Toda potential for the associated Lie algebra~\cite{Davies:2000nw}
\begin{equation}
  \label{eq:W-BPS}
  W(\Sigma)_{\text{BPS}} \equiv \sum_{i=1}^{\rank(G)} \frac{2}{\alpha_i^2} \exp[\alpha_i^* \cdot \Sigma] \; ,
\end{equation}

\item The picture incorporates very naturally the \acs{kk} monopoles due to the compact direction.
  It turns out that the extra \F1 string, which winds around the circle connecting, for unitary $G$, the $1$st and the $N$th \D3 brane,
  can be accounted for by extending the Dynkin diagram to its \emph{affine} version.
  We depicted this in Figure \ref{fig:Dynkin-An}.

  Summing the contributions from the \acs{bps} and the \acs{kk} monopoles we obtain the superpotential on the compact Coulomb branch,
  finding an affine Toda potential  for the associated affine algebra~\cite{Davies:2000nw}
\begin{equation}
  \label{eq:W-afffine}
  W(\Sigma) = W(\Sigma)_{\text{BPS}} + W(\Sigma)_{\text{KK}} \equiv
  \sum_{i=1}^{\rank(G)} \frac{2}{\alpha_i^2} \exp[\alpha_i^* \cdot \Sigma] +  \frac{2 \eta}{\alpha_0^2}
\exp[\alpha_0^* \cdot \Sigma] \; ,
\end{equation}
where $\Sigma= \sigma/e_3^2+ i  \phi$, $\alpha_i$ are the simple roots and $\alpha^*_i$ are the associated co-roots.
The extra simple root $\alpha_0$ corresponds to the \ac{kk} monopole
and the corresponding contribution $W(\Sigma)_{\text{KK}}$ to $\eqref{eq:W-afffine}$ is identified with $W_\eta$ in field theory.

\item
  After this general recipe let us spell out some details for the example of unitary gauge groups. Here 
  we have the affine algebra  $\widetilde A_N$, where the extra simple root is associated to
the combination $\sigma_N-\sigma_1$ (see Figure~\ref{fig:Dynkin-An}).
\begin{figure}
  \centering
  \includegraphics[width=14cm]{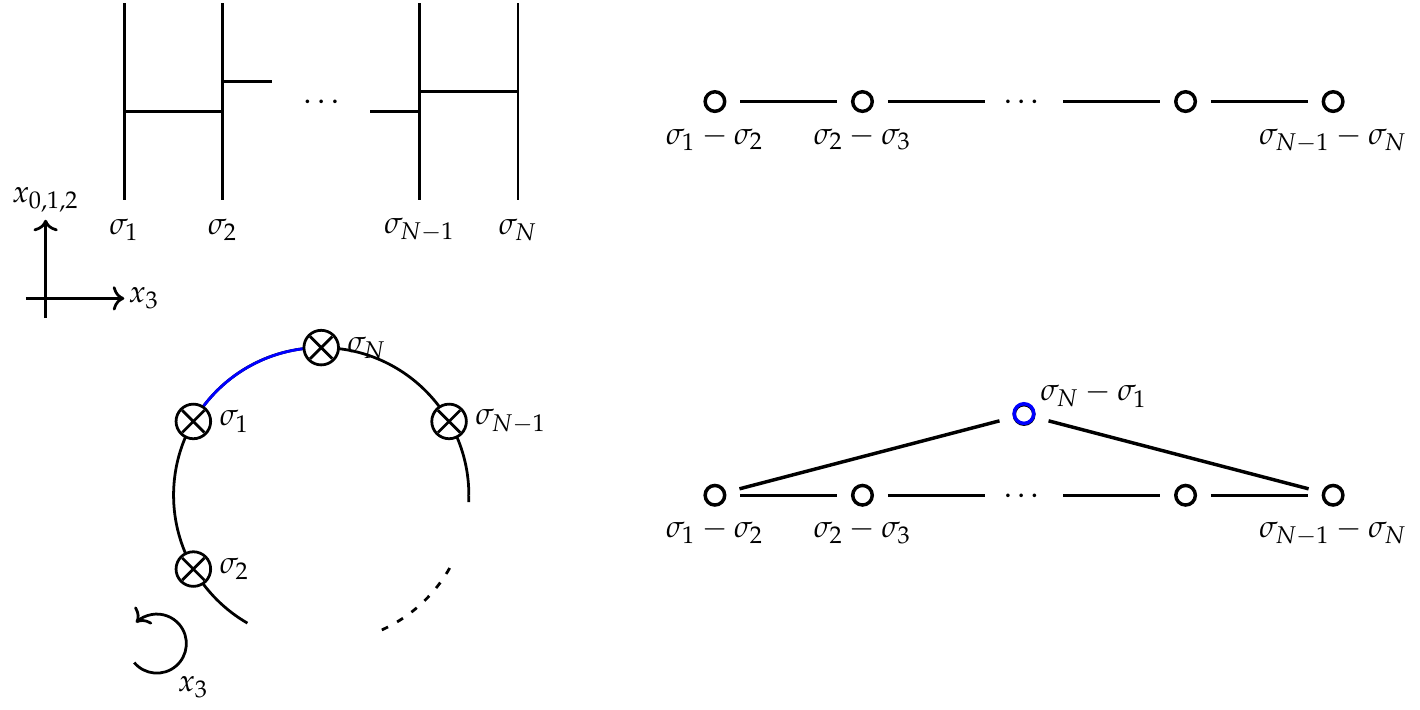}
   \caption{Branes and Dynkin diagrams for \(A_N\) and \(\widetilde
    A_N\).  The left column shows the S-dual configuration of \F1--strings stretched between \D3--branes.
  In the right column we depict the corresponding \(A_N\) and
  \(\widetilde A_N\) Dynkin diagrams.
  After compactification a new string appears between
  \(\sigma_1\) and \(\sigma_N\) and corresponds to the affine node
  (in blue in the brane cartoon and in the Dynkin diagram).}
  \label{fig:Dynkin-An}
\end{figure}
For $SU(N)$ theories\footnote{In the $U(N)$ case the same result holds but the last (affine) root splits in two terms
$Y_{+} =e^{\sigma_1/e_3^2+i\phi_1}$ and  $Y_{-}=e^{-\sigma_{N}/e_3^2-i \phi_N}$.} the superpotential $\eqref{eq:W-afffine}$,
associated to the $\widetilde A_N$ diagram, is
\begin{equation}
\label{etaWSU}
W = \sum_{i=1}^{N-1} \frac{1}{Y_i}  + \eta Y_N \; ,
\end{equation}
where $Y_i =  e^{\Sigma_i-\Sigma_{i+1}}$.
The last term in (\ref{etaWSU}) breaks explicitly the $U(1)_A$ symmetry
in the three-dimensional field theory. This symmetry is associated to the rotation
in the $(4,5)$ plane in the brane picture. The geometric realization
of the breaking of this symmetry for compact $x_3$ has been discussed in~\cite{Amariti:2015yea}.
\item
Next we want to discuss real gauge groups.
Here we outline few aspects, the proper analysis is given in the next sections.
They are realized by including \O3 or \O5 planes.
Let us first discuss the configurations with \O3 planes on $\mathbb{R}^3 \times S^1$.
As reviewed in Appendix \ref{sec:orientifolds} the four differently charged orientifolds
\O3$^+$, \O3$^-$, $\widetilde \text{\O3}^+$ and $\widetilde \text{\O3}^-$
project a unitary group to $SP(2N), SO(2N), SP(2N)$ and $SO(2N+1)$ respectively.
In presence of a compact direction orientifolds come in pairs and
here\footnote{For simplicity we do not distinguish between \O3$^+$ and $\widetilde \text{\O3}^+$ planes.
} we have six different such pairs~\cite{Hanany:2001iy}.
Brane configurations with $(\O3^-,\O3^-)$,  $(\widetilde \O3^-,\O3^-)$ and $(\O3^+,\O3^+)$
are associated to affine Dynkin diagrams.
The other pairs $(\O3^{+},\O3^{-})$, $(\O3^{+},\widetilde \O3^{-})$ and $(\widetilde \O3^{-}, \widetilde \O3^{-})$ correspond to twisted affine Dynkin diagrams.
In this paper we are interested only in the ``affine'' pairs,
since those are obtained by a T--duality from \tIIA configurations with \O4--planes.

A similar discussion holds with $\O6$--planes, while the effect of the orientifold charge on the projection is exchanged.
$\O5^+$ is associated to an $SO(N)$  and $\O5^-$ to an $Sp(2N)$ gauge group.
The pairs $(\O5^\pm,\O5^\pm)$ are obtained by T--duality from a \tIIA configuration with $\O6^{\pm}$ planes.
\end{itemize}

\bigskip

Eventually we include matter fields.
Standardly, 
we can couple them to the four dimensional gauge theory by adding stacks of \D6--branes in the \tIIA setup,
as shown in Table~\ref{tab:O4+_branes}.
In the T--dual frame they become \D5--branes.

\bigskip

So far the brane configurations for the dynamics of gauge theories on $\mathbb{R}^3 \times S^1$. 
Let us now turn to dualities and how their dimensional reduction can be understood in this picture.
Here we highlight the steps following the example of $U(N)$ gauge theories with fundamental matter,
in the next sections we apply them to more general cases.

The brane construction of Seiberg dualities in $4$D is well understood,
it boils down to an \ac{hw} transition in the \tIIA configuration.
In the \ac{hw} transition every time a \D6--brane crosses a non-parallel fivebrane, a \D4--brane is generated.
This brane creation mechanism in necessary for charge conservation, to preserve the so called linking number~\cite{Hanany:1996ie}.
Moving the entire stack of flavor \D6--branes from one side of the \NS~to the other
amounts to changing the brane configuration of the electric gauge theory to the configuration describing the magnetic one. 

For example unitary \textsc{sqcd} with $F+k$ flavors, \emph{i.e.\ } $F+k$ pairs of fields in the fundamental and anti-fundamental representation of $U(N)$,
is engineered by a stack of $N$ \D4--branes stretched between an \NS~and an \NS', with $F+k$ \D6--branes on top as denoted in Table~\ref{tab:O4+_branes}. 
The magnetic dual is obtained by swapping the \NS~with the \NS',
changing the number of \D4--branes in the stack to $F+k-N$.

From the $4$D brane picture we can obtain the one describing Seiberg duality on $\mathbb{R}^3 \times S^1$ by a T--duality
as described earlier in this section.
The T--dual of the configuration describing the $4$D electric theory 
maps to the brane configuration of the effective $3$D electric theory with $W_\eta$ 
and analogously the T--dual of the $4$D magnetic brane configuration maps to the brane setup of the magnetic theory with $W_\eta$.

In our example of unitary \textsc{sqcd} we obtain the \tIIB configuration with a stack of $N$ \D3 and $F+k$ \D5--branes at finite radius of $x_3$.
This brane setup describes the electric theory with $U(N)$ gauge group, the superpotential $W_\eta$, and $F+k$ flavors.
The magnetic dual configuration has $F+k-N$ \D3 and $F+k$ \D5--branes,
it describes a $U(F+k-N)$ gauge theory with the superpotential $W_\eta$, $F+k$ flavors and $(F+k)^2$ gauge singlets.

\bigskip
Ultimately we want to obtain pure $3$D dualities in this brane picture.
In order to reproduce the construction in~\cite{Aharony:2013dha}
we move, as depicted in Figure~\ref{fig:new-orientifold},
some flavor branes of the electric theory to the mirror point of the T--dual circle. 
In field theory this corresponds to giving a mass $\sim \mathcal{O}(\frac{\alpha'}{r})$ to the associated matter fields.
When taking the radius $r$ to zero, it corresponds to the double-scaling limit mentioned in the introduction and used in the next sections. 

The magnetic dual can be obtained by an \ac{hw} transition swapping the \NS--branes in the \tIIB configuration. 
This is equivalent to a double-scaling limit in the magnetic theory, that we had obtained by T--duality from \tIIA,
when the flavor \D5--branes each drag a \D3 to the mirror point. 
In field theory it hence reproduces the higgsing of the theory.

More explicitly, in the example of unitary \textsc{sqcd},
we move one stack of $k$ \D5--branes clockwise and another one counterclockwise on the circle,
until they reconnect at the mirror point $x^\circ_3$.
When taking the limit $r \rightarrow 0$ the \D5--branes at $x^\circ_3$ correspond to a set of massive fields,
which do not contribute to the low energy theory.
On the other hand in the magnetic theory, the \D5--branes at $x^\circ_3$ do give rise to massless states, contributing to the low energy dynamics.
The reason is that the \ac{hw} transition creates $k$ \D3--branes at $x^\circ_3$. 
In terms of field theory, this corresponds to an extra $U(k)$ gauge theory with $k$ massless fundamental flavors and $k^2$ massless singlets.
The singlets interact with the flavor fields through a superpotential, as the \D5 and the \NS' are parallel.

In all cases studied in this paper the extra gauge sector in the magnetic theory at $x_3 = x_3^\circ$ can be described as a theory of interacting gauge singlets.
Furthermore, this extra sector always interacts with the monopoles of the magnetic theory at $x_3 = 0$ and 
the interaction involves some of the gauge singlets describing the theory at $x_3 = x_3^\circ$.
Next we discuss the example of unitary gauge groups with fundamental flavor, other examples will be described in the next sections.

In the example of \textsc{sqcd} the extra sector corresponds to a $U(2k)$ gauge theory with $2k$ flavors $f$ and $\tilde f$,
a singlet $L$ with $4k^2$ components and the superpotential $W=L f \tilde f$.

This theory is ``mirror'' dual~\cite{Aharony:1997bx} to a set of singlets $M$, $L$ and $V_\pm$, 
where $M$ is identified with the meson $M \equiv f \tilde f$ and the singlets $V_\pm$ are identified with the monopoles of the $U(2k)$ sector,
$V_+ \equiv e^{\Sigma_1}$ and $V_- \equiv e^{-\Sigma_{2k}}$, where $\Sigma_i$ is the $i$th \textsc{cb} coordinate of the $U(2k)$ gauge theory.
The mirror theory is interacting, with superpotential $W=L M + V_+ V_- \det M$.
In the \textsc{ir} this superpotential is set to zero by the eom.

We just have argued that the $U(2k)$ gauge theory at $x_3=x_3^\circ$ is effectively described by a theory of interacting gauge singlets. 
Now we want to come back to the brane picture of $3$D duality. 
There is an interaction between this $U(2k)$ sector and the magnetic $U(F-N)$ gauge theory at $x_3 = 0$.
This interaction can be studied, as in the beginning of this section, by \D1--branes stretching between the two stacks of \D3--branes.
More explicitly, they describe a repulsive force between the $1$st \D3 brane at $x_3 = 0$ and the $(2k)$th \D3--brane at $x_3^\circ$,
and analogously a repulsive force between the $(F-N)$th \D3--brane at $x_3=0$ and the $1$st \D3-brane at $x_3^\circ$.
In field theory language this force is manifest through the superpotential\footnote{This 
superpotential corresponds to an \ac{ahw} superpotential due to the higgsing of the magnetic gauge theory.}
\begin{align}
W= y_+ V_- + y_- V_+ \, ,
\label{eq:W-Aharony}
\end{align}
where $y_+ = e^{\Sigma_1}$ and $y_- = e^{-\Sigma_{F-N}}$ and $\Sigma_i$ is the $i$th \textsc{cb} coordinate of the $U(F-N)$ gauge theory at $x_3 =0$.

Note that the superpotential $\eqref{eq:W-Aharony}$ survives in the mirror dual descrition of the $U(2k)$ gauge theory.
Indeed, the monopoles of the $U(2k)$ sector, which interact with the monopoles of the $U(F-N)$ sector, are exactlty those which are identified with the singlets 
$V_+$ and $V_-$ under mirror symmetry.

The interaction $\eqref{eq:W-Aharony}$ can be seen as generating the relations $y_\pm = 0$ on the chiral ring of the magnetic theory with gauge group $U(F-N)$,
as in the Aharony duality.

Indeed, given their interaction with the monopoles $y_\pm$ of the magnetic theory,
the singlets $V_+$ and $V_-$ have the natural interpretation as monopoles of the electric theory.
In this sense we have recovered a brane description of the dynamics of Aharony duality.

\bigskip

In the rest of this paper we illustrate the generality of this picture, considering the effects of orientifold planes.
Their \D-brane charge modifies the standard brane creation effect in the \ac{hw} transition.
Nevertheless, the extra sector at $x^{\circ}_3$ in the magnetic theory remains mirror dual to singlets.

\section{Sp(2N) theories}

\label{sec:symplectic}

In this section we discuss the reduction of the duality for $Sp(2N)$
with $2F$ fundamentals.
An  $Sp(2N)$ gauge theory with $2 F$ fundamentals%
\footnote{In some cases the flavor symmetry
is $SU(F)^2$ instead of $SU(2F)$ and we have $F$
pairs of fundamental and anti fundamental. We will
denote this possibility as having $F$ flavors.} and
global symmetry $SU(2F) \times U(1)_A \times U(1)_R $
without superpotential is dual to
an $Sp(2(F-N-2))$ gauge theory with $2F$ dual fundamentals $q$ and a meson $M$
with superpotential $W= M q q $.
This duality was first presented in~\cite{Intriligator:1995ne}.
The $U(1)_A$ symmetry is anomalous at the quantum level. We
present the global charges associated to this symmetry
because it is quantum realized in the three-dimensional case.
The field content is given in Table~\ref{tab:field-content-Sp}.
\begin{table}
  \centering
  \begin{tabular}{ccccccc}
    \toprule
          & \(Sp(2N)\) & \(Sp(2 \widetilde N)\) & \(SU(2F)\)         & \(U(1)_A\) & \(U(1)_R\)     \\ \midrule
    \(Q\) & \(2N\)     & \(1\)                  & \(2F\)             & \(1\)      & \(1-(N+2)/F\)  \\
    \(q\) & \(1\)      & \(2 {\widetilde N}\)   & \( 2 \overline F\) & \(-1\)     & \((N+2)/F\)    \\
    \(M\) & \(1\)      & \(1\)                  & \(F(2F-1)\)        & \(2\)      & \(2-2(N+2)/F\) \\ \bottomrule
  \end{tabular}
  \caption{Field content for the \( Sp(2N) \) gauge theory with  global \( SU(2F) \times U(1)_A \times U(1)_R\) symmetry.}
  \label{tab:field-content-Sp}
\end{table}

\subsection{Brane description}

There are two ways to represent this theory, by either considering an $\O4^{+}$--plane
or an $\O6^{-}$--plane.
These two constructions give rise to similar theories, that differ for the representation
of the matter fields under the global symmetries.
When modifying the theory (by
allowing larger numbers of \NS{}--branes) the two constructions give rise to different
two-index matter fields, adjoint or antisymmetric. We will study both possibilities.
\begin{itemize}
\item In the $\O4^{+} $--plane case, the brane setup is summarized in~Table~\ref{tab:O4+_branes} and Figure~\ref{fig:O4+_branes}.
In this case we consider a stack of \(2N\) \D4--branes and an  $\O4^{+} $--plane
stretched between an \NS{} and an \NS'--brane. We consider also \(2F\) \D6--branes
on the \NS'--brane.
At brane level this theory as an $SO(2F)$ global symmetry, while on the field theory side there is the enhancement to
$SU(2F)$. This is similar to the usual doubling of the global symmetry for unitary gauge groups. 
The dual theory is obtained by an \ac{hw} transition that exchanges the \NS{} and the \NS'--branes.
In presence of an orientifold the linking number is modified. For example
an $\O4^{+}$ has to be treated as a stack of $-4$
\D6--branes~\cite{Elitzur:1997hc}.
After the transition we obtain the dual picture, in which the net number of \D4--branes is
$2(F-N-2)$.
\begin{figure}
  \centering
  \includegraphics{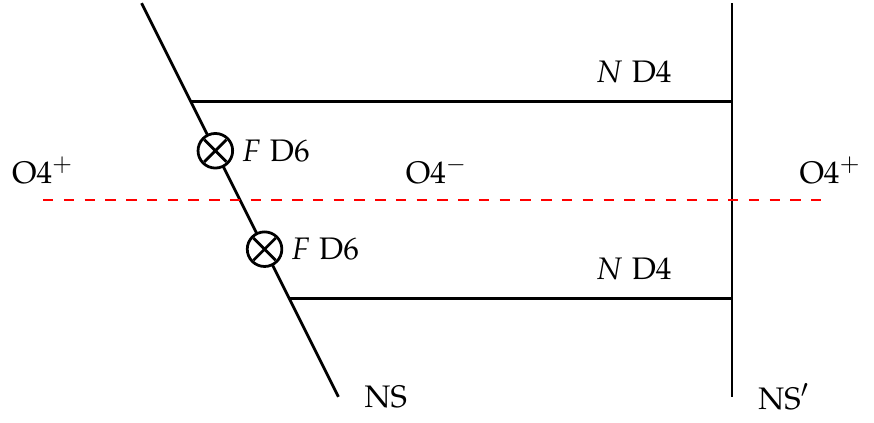}
  \caption{Brane cartoon for the realization of an \(Sp(2N)\) theory in the electric phase with an \O4--plane.}
  \label{fig:O4+_branes}
\end{figure}
\item A similar theory can be constructed by using an $\O6^{-}$--plane.
Consider two \NS--branes, $2N$ \D4s, $2F$ \D6's  and an
$\O6^{-}$--plane as in Figure~\ref{fig:O6-_plane_duality_4D}(a). If
all the \NS--branes are parallel, the system has $\mathcal{N}=2$ supersymmetry.
The orientifold projects the $SU(2N)$ gauge group to $Sp(2N)$
(where, as usual, $Sp(2)\simeq SU(2)$).
The theory has an $SU(F)$
global symmetry with $F$ flavors. We expect this symmetry to be enhanced to $SU(F)^2$.
Here we rotate the \NS--branes and the \D6'--branes by an angle $\theta$ as in
Figure~\ref{fig:O6-_plane_duality_4D}(a), and we have two stacks of $\NS_{\pm \theta}$ and $\D6_{\pm \theta}$.
For generic angles the $\mathcal{N}=2$
adjoint is massive. If $\theta=\pi/2$ the orientifold is parallel
to the $NS_{\pm \theta}$--branes and this field is massless and has to be considered in the low-energy spectrum. We will come back to this configuration later.
This model (for $\theta \neq \pi/2$) has a dual description as discussed above. In this case the
$\O6^{-}$ behaves like a stack of $- 4$  \D6--branes in the \ac{hw} transition. The brane picture becomes the one shown in Figure~\ref{fig:O6-_plane_duality_4D}(b) where the dual gauge group is again $Sp(2(F-N-2))$.
\begin{figure}
  \centering
  \includegraphics{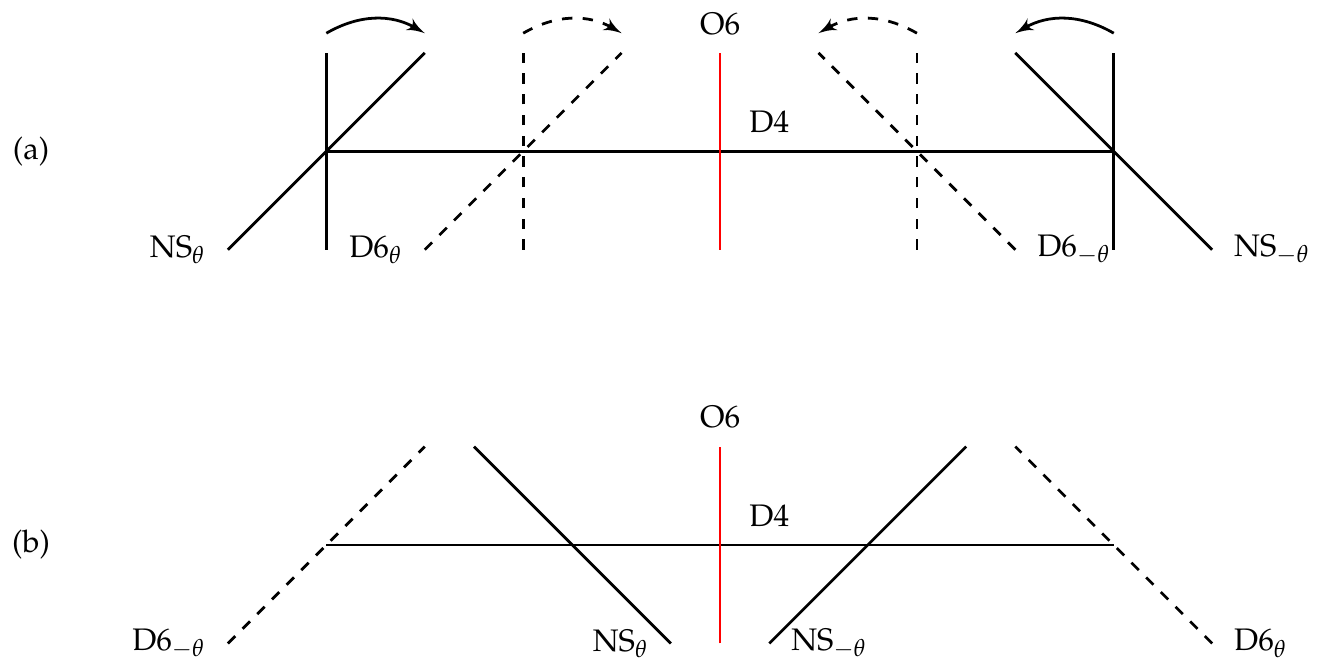}
  \caption{Brane cartoon for the realization of an \(Sp(2N)\) theory in the (a) electric and (b) magnetic phase with an \O6--plane.}
  \label{fig:O6-_plane_duality_4D}
\end{figure}
\end{itemize}

\subsection{Dimensional reduction}

\subsubsection{\O3--planes}
Let us begin with the reduction of the duality  with an $\O4^{+}$--plane.
The three-dimensional system is obtained by compactifying the $x_3$--direction and T--dualizing. The \NS{}--branes remain invariant while the orientifold
becomes an pair of ($\O3^+,\O3^+$)--planes.
We can study the properties of the Coulomb branch by looking at the spectrum of \acs{bps} \F1 strings as
explained above.
\begin{figure}
  \centering
  \includegraphics{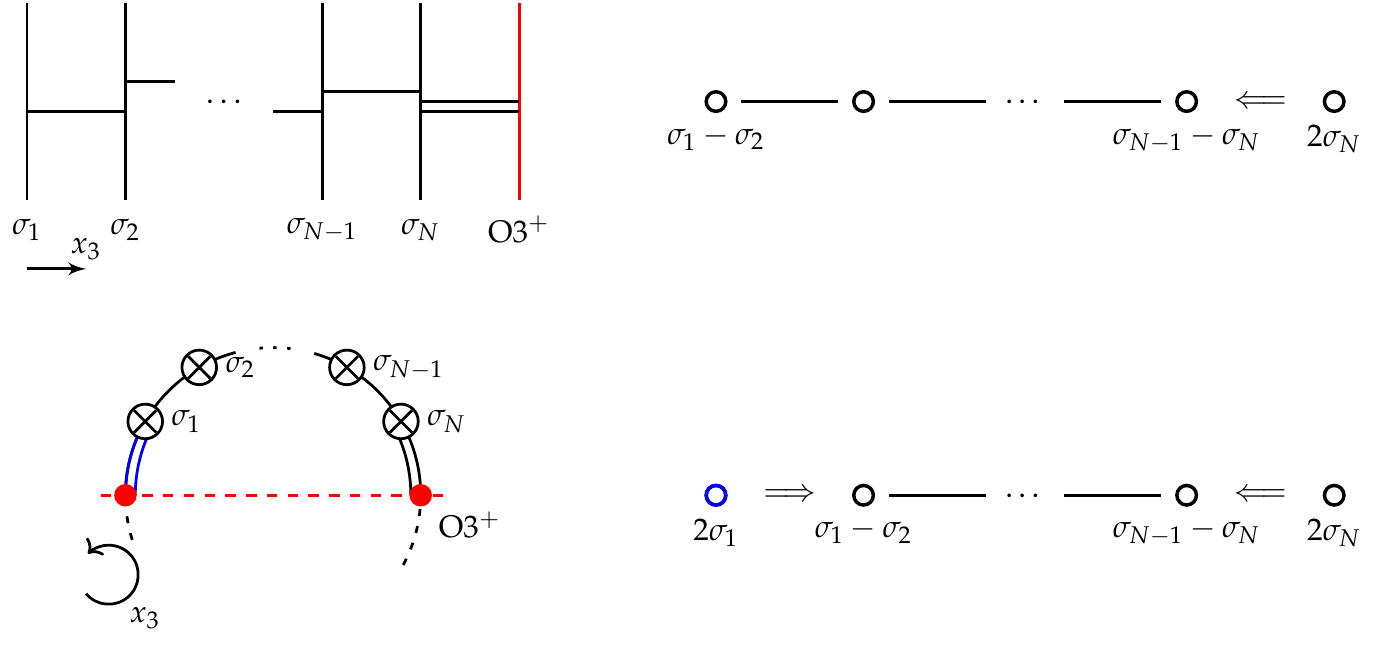}
  \caption{Dynkin and affine Dynkin diagrams and spectrum of \acs{bps} \F1--strings
  associated to the fundamental monopoles for $Sp(2N)$ theories
   in the linear case and on the circle. The affine root is represented in blue on the affine 
   Dynkin diagram and in the brane cartoon.
   }
  \label{fig:F1Cn}
\end{figure}
In this case the superpotential can be read off from the top half of Figure~\ref{fig:F1Cn},
\begin{equation}
  W = \sum_{i=1}^{N-1} \frac{2}{Y_i} + \frac{1}{Y_N} \; ,
\end{equation}
where $Y_i = e^{(\sigma_i-\sigma_{i+1})/e_3^2+i(\phi_i-\phi_{i+1})}$
and $Y_N=e^{2(\sigma_N/e_3^2+i \phi_N)}$.
The extra root in the affine case is proportional to the variable $Y_0 = e^{2(\sigma_1/e_3^2+i \phi_1)}$ (shown in blue on
the botton half of Figure~\ref{fig:F1Cn}) and it gives the superpotential
\begin{equation}
  W_\eta = \eta \eu^{2 \sigma_1/{e_3^2} + 2 i \phi_1} \; .
\end{equation}
The same result is obtained after the \ac{hw} transition.

\bigskip

Now we want to flow to the Aharony duality.
We start by considering $2(F+1)$ \D5--branes in the electric theory.
We rotate two \D5--branes on the circle and reconnect them
on the other side of the circle.
Since the \D5s intersect the \NS{}--brane in this configuration,
there are no massless fields in this extra sector.
If we take the $r \rightarrow 0$ limit on this configuration, we obtain
an $Sp(2N)$ theory with $2F$ fundamentals.

Next we turn to the dual theory. In this case if we perform an \ac{hw} transition,
there are $2(F-N-1)$ \D3s at the origin. On the other side of the circle the two \D3s created
by the \D5 crossing the \NS{}--brane are destroyed by the extra orientifold plane located there.
This example shows one of the general aspects of our analysis. 
In principle it is not necessary to reconnect the \D5--branes at $x_3^\circ$.
For example for unitary gauge groups the double scaling was realized by putting the \D5--branes at $x_3 < x_3^\circ$~\cite{Amariti:2015yea}.
Here avoiding the orientifold to create a negative number of \D3--branes in the \ac{hw} transition we have to reconnect the \D5s at $x_3^\circ$.
In the rest of the paper we will always follow this strategy.

\begin{figure}
  \centering
  \includegraphics[width=12cm]{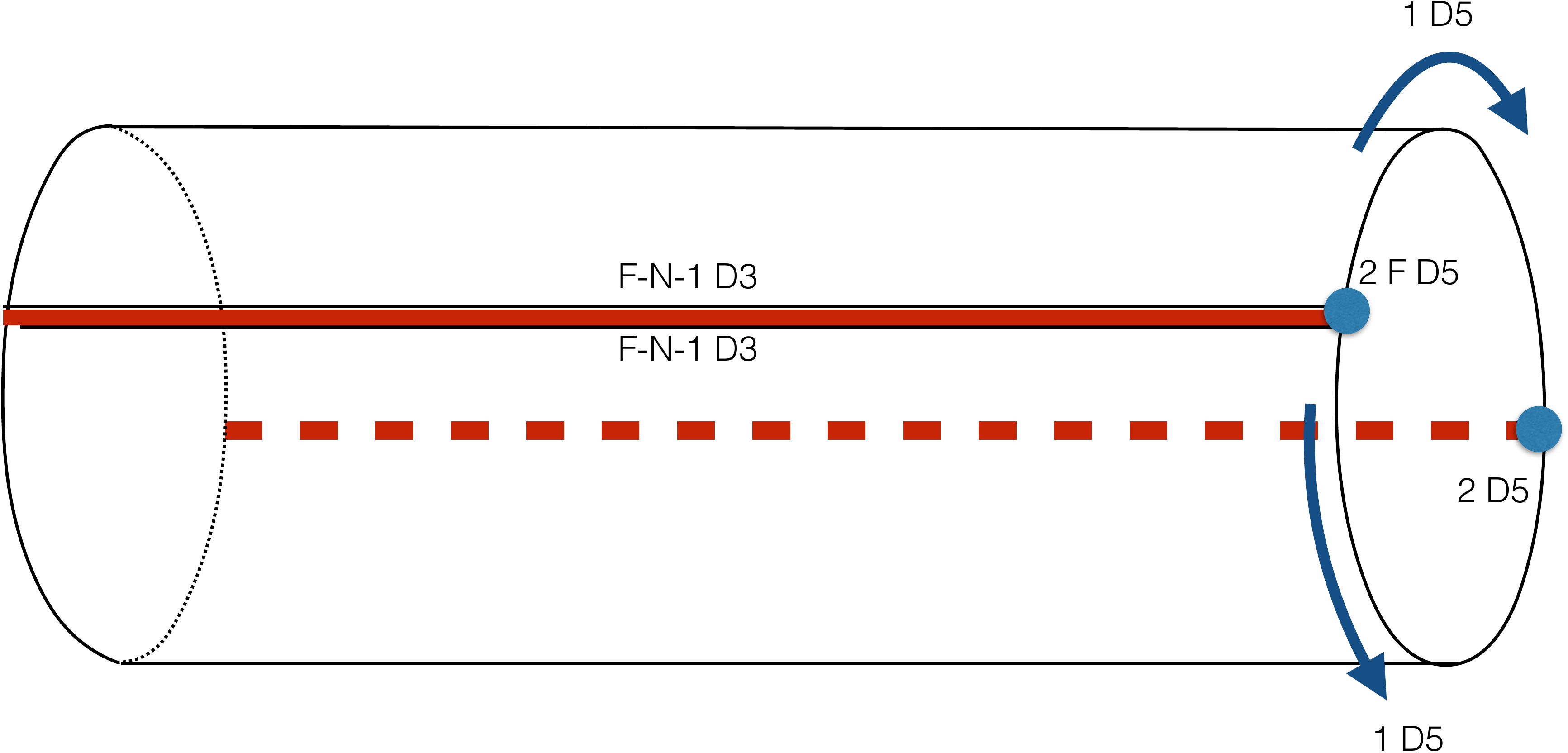}
  \caption{Dual Aharony flow \(Sp\) in the \O3 configuration}
  \label{fig:Aharony-flow-Sp-O3}
\end{figure}

The final configuration is represented in Figure~\ref{fig:Aharony-flow-Sp-O3}.
In this case, even if there is no gauge symmetry, an extra meson arising from the \D5--brane remains massless. This suggests that we cannot simply decouple this sector before considering the effect of this massless field in the ordinary dual gauge theory.
In fact in this case the two \D5--branes attract the branes labeled by $\sigma_1$ and
$-\sigma_1$ in the dual gauge sector.
This attractive force is reflected in the scale-matching relation between $Y_1$ and
the meson $M_{2F+1,2F+2}$.
It corresponds to the superpotential interaction
\begin{equation}
W = \widetilde \eta y_{low} M_{2F+1,2F+2} \; ,
\end{equation}
\emph{i.e.} the low-energy description of the superpotential $W_\eta$.
In the large-mass limit
the effect of this interaction has to be considered.

This reproduces the field theory expectation:
the dual theory is an $Sp(2(F-N-1))$ theory
with $2F$ fundamentals, an antisymmetric
meson $M$ and superpotential
\begin{equation}
W = M q q + y Y \; ,
\end{equation}
where we identified the broken component of the electric singlet \(M\) that parametrizes
a direction in the dual Higgs branch, with the electric monopole \(Y\) that parameterizes the Coulomb branch of the electric phase. This is commonly the case when dealing with mirror symmetry and in fact the electric singlet
describes the Higgs branch of the dual phase, \emph{i.e.} the Coulomb branch of the
electric theory.

\subsubsection{\O5--planes}

Also in the case of the \O6--plane realization one can reduce the duality to three dimensions by compactifying the $x_3$--direction. After T--duality the \tIIB system contains a pair of \O5--planes and describes a theory with the same superpotential $W_\eta$ as above. By considering $F+2$ flavors and by integrating out of them we recover the usual Aharony duality. At the brane level
this is obtained by introducing $F+2$ $\D5_{\pm \theta}$.
We introduce real masses as in the construction with the \O3--plane. The orientifold identification is however different: in this case
we have a unitary symmetry. Moving a pair $\D5_{\pm \theta}$ along $x_3$ gives a mass to one flavor.

 \begin{figure}
  \centering
  \includegraphics[width=.8\textwidth]{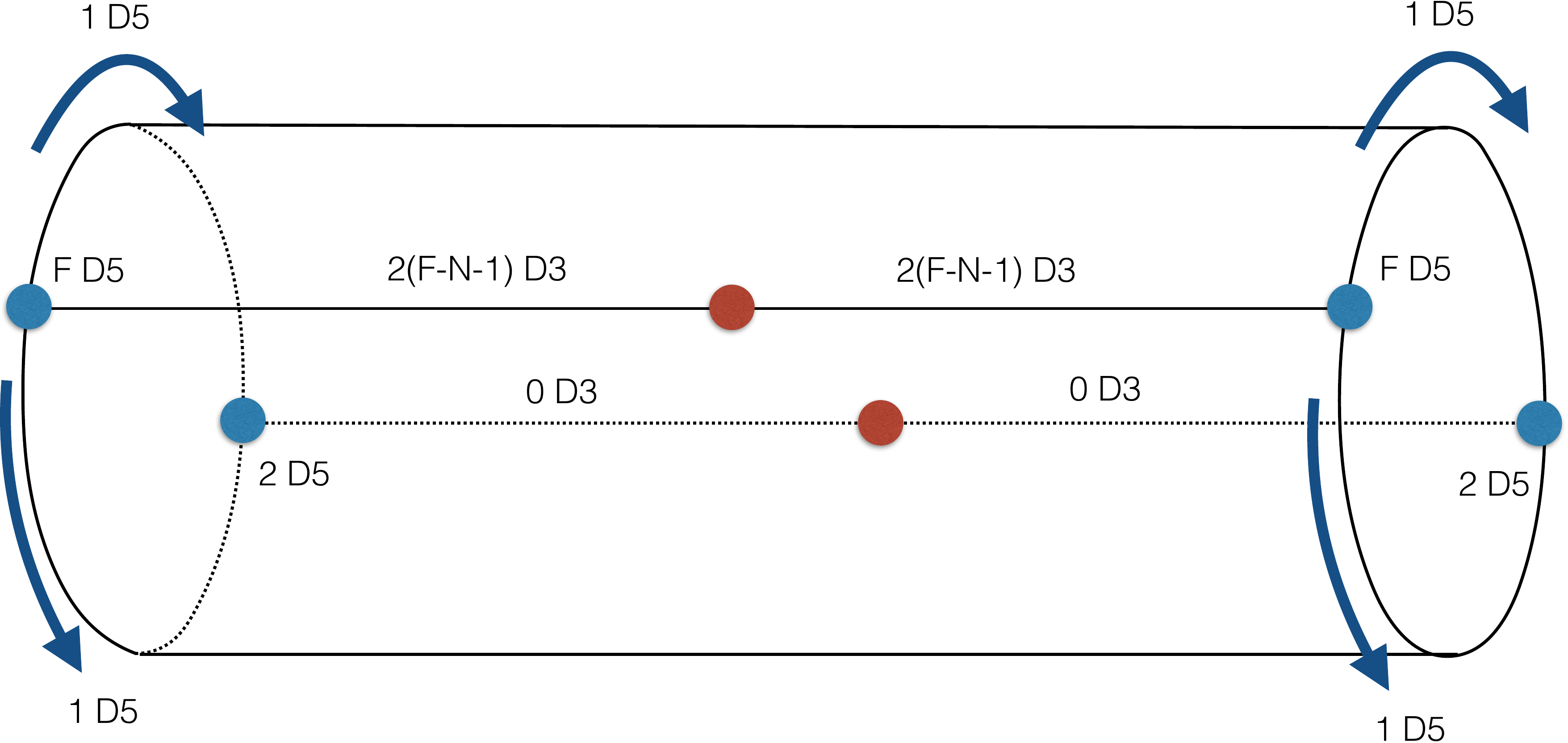}
  \caption{Aharony flow \(Sp\) in \O5 picture}
  \label{fig:Aharony-flow-Sp-O5}
\end{figure}

One can flow to Aharony duality by taking the double scaling limit as above (see Figure~\ref{fig:Aharony-flow-Sp-O5}).
Let us explain the duality in this case.
First we move the \D5--branes in the $x_3$--direction, assigning the real masses.
We reconnect them on the other side of the circle. They reconnect at
$x_3 = x_3^\circ$, where the second orientifold plane is located.
The extra sector does not have massless degrees of freedom, and we can take the
$r \rightarrow 0$ limit in this case. We obtain a three-dimensional $Sp(2N)$ theory
with $2F$ flavors.
Now we can turn to the dual picture, by exchanging the $\NS{}_{\pm \theta}$--branes.
The \D3s are created when the branes cross each other.
While at the origin the orientifold cancels two \D3s every time an \NS--brane crosses it,
the net effect on the \D3s at  $x_3 = x_3^\circ $ is the absence of branes in the gauge theory.
The final configuration is reproduced in Figure~\ref{fig:Aharony-flow-Sp-O5}.

Like in the case with \O3--planes, here we have an extra sector with massless singlets (coming from the original mesons). The $r \rightarrow 0$ limit has to be taken by considering the effect of this sector on the  $Sp(2(F-N-1))$ theory. This is the same mechanism introduced above: the superpotential $W_\eta$ is absorbed in a scale matching, the meson couples with the magnetic monopoles, and in the final three-dimensional dual theory the extra interaction between the electric and magnetic monopoles takes place.

%
%
%
%
%
%

\subsection{Generalizations}

\subsubsection{Sp(2N) with antisymmetric matter}

An $Sp(2N)$ gauge theory with $F$ flavors $Q$ and $\widetilde Q$ and an antisymmetric field $A$,
 with superpotential \(W = \Tr A^{k+1}\) is dual to
an $Sp(2(k(F-2)-N))$ gauge theory with $F$ flavors $q$ and $\tilde q$, an antisymmetric $a$ and superpotential \(W = \Tr a^{k+1} + \sum_{j=0}^{k-1} M_{k-j-1} q a^j \tilde q\), where $M_j = Q A^{j} \widetilde Q$ is the generalized meson, with $j=0,\dots,k-1$.
This duality was first presented in~\cite{Intriligator:1995ff}.

\begin{figure}
  \centering
  \includegraphics{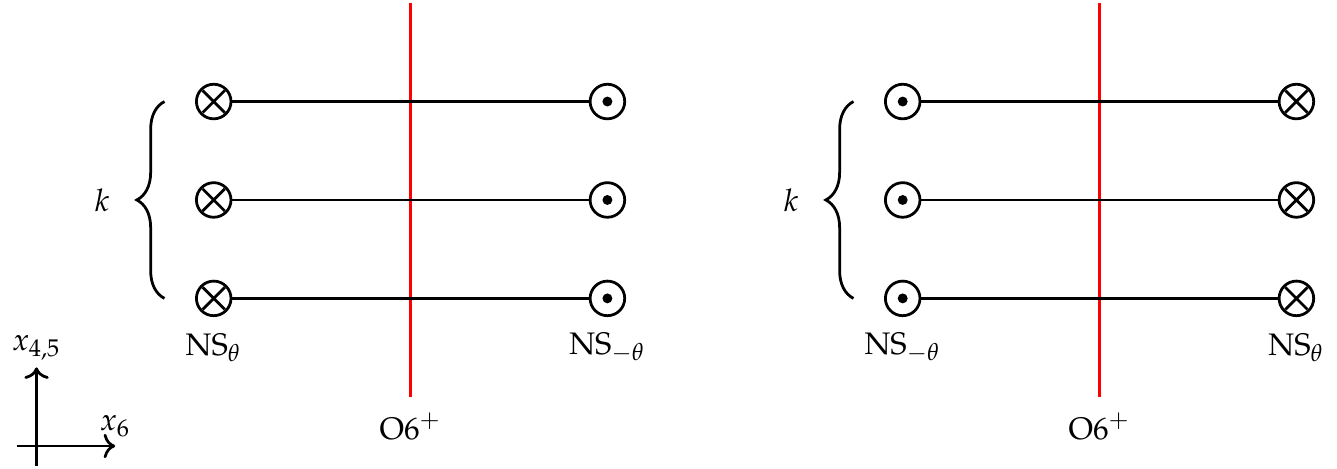}
  \caption{Electric and magnetic sides of the duality for \(Sp(2N)\) gauge theories with antisymmetric matter.}
  \label{fig:Sp-antisymmetric}
\end{figure}

In this case we consider two stacks of $k$ $\NS{}_{\pm \theta}$--branes.
The gauge symmetry is broken by
separating them along the directions \(4\) and \(5\) and leads to a polynomial superpotential
for the antisymmetric field $A$.
The electric theory is broken to
\begin{equation}
Sp(2 N) \rightarrow \prod_{i=1}^{k} Sp(2 r_i) \;,
\end{equation}
while the magnetic one becomes
\begin{equation}
Sp(2 \widetilde N) \rightarrow \prod_{i=1}^{k} Sp(2 \widetilde r_i) \;,
\end{equation}
where $\widetilde r_i = F- r_i - 2$ (see Figure~\ref{fig:Sp-antisymmetric}).
%

In this case we can perform the reduction on each sector. The bare monopoles
associated to each $Sp(2 r_i)$ factor recombine, through the scale matching relation, with the
antisymmetric field when the superpotential deformations are turned off. This
correspond to recombining the $NS_{\pm \theta}$--branes.

The theory on the circle can be further reduced to an Aharony-like duality by integrating
out some matter fields. In the three-dimensional case we refer to the antisymmetric
representation discussed in~\cite{Kapustin:2011vz}, obtained by combining the irreducible antisymmetric with a singlet.

If we consider $(F+2)$ flavors and integrate out two of them in each sector
we arrive in the dual at a
$ \prod_{i=1}^{k} Sp(2 (F-r_i-1)) $
gauge theory. After reconnecting the branes, the dual theory is
$ Sp(2 (k(F-1)-N)) $.
As a check we consider $(F+2K)$ flavors and flow to a known duality.
Integrating out $2K$ flavors after assigning them the same large real mass generates a \ac{cs} term.
We arrive at the duality of Kapustin, Kim and Park~\cite{Kapustin:2011vz}
between\footnote{The \ac{cs} levels have an extra factor of two because of the
normalization of the generators in the Lie algebra.}
$Sp(2N)_{2K}$ and  $Sp(2 (k(F+|K|-1)-N))_{-2K}$.

\subsubsection{Sp(2N) with adjoint matter}

An $Sp(2N)$ gauge theory with $2F$ fundamentals
and an adjoint field $X$, with superpotential \( W = \Tr (X)^{2(k+1)}\)
is dual to
an $Sp(2((2k+1) F-N-2))$ gauge theory with $2F$ fundamentals, an adjoint $Y$ and superpotential
\begin{equation}
\label{eq:dualadjsp}
W = Y^{2(k+1)} +
\sum_{j=0}^{2k} M_{2k-j} q Y^j q \;,
\end{equation}
where $Y$ is in the adjoint of the dual group and $M_j = Q X^j  Q$.
This duality was first presented in~\cite{Leigh:1995qp}.

The electric theory is represented by $2 N$ \D4--branes and an $\O4^-$--plane stretched between $2k+1$ \NS{}--branes and one \NS'.  In addition, there are $2F$ \D6--branes on the \NS{}--branes.
By separating the \NS{}--branes along the $(45)$--plane we have a polynomial deformation in the adjoint $X$.

\begin{figure}
  \centering
  \includegraphics{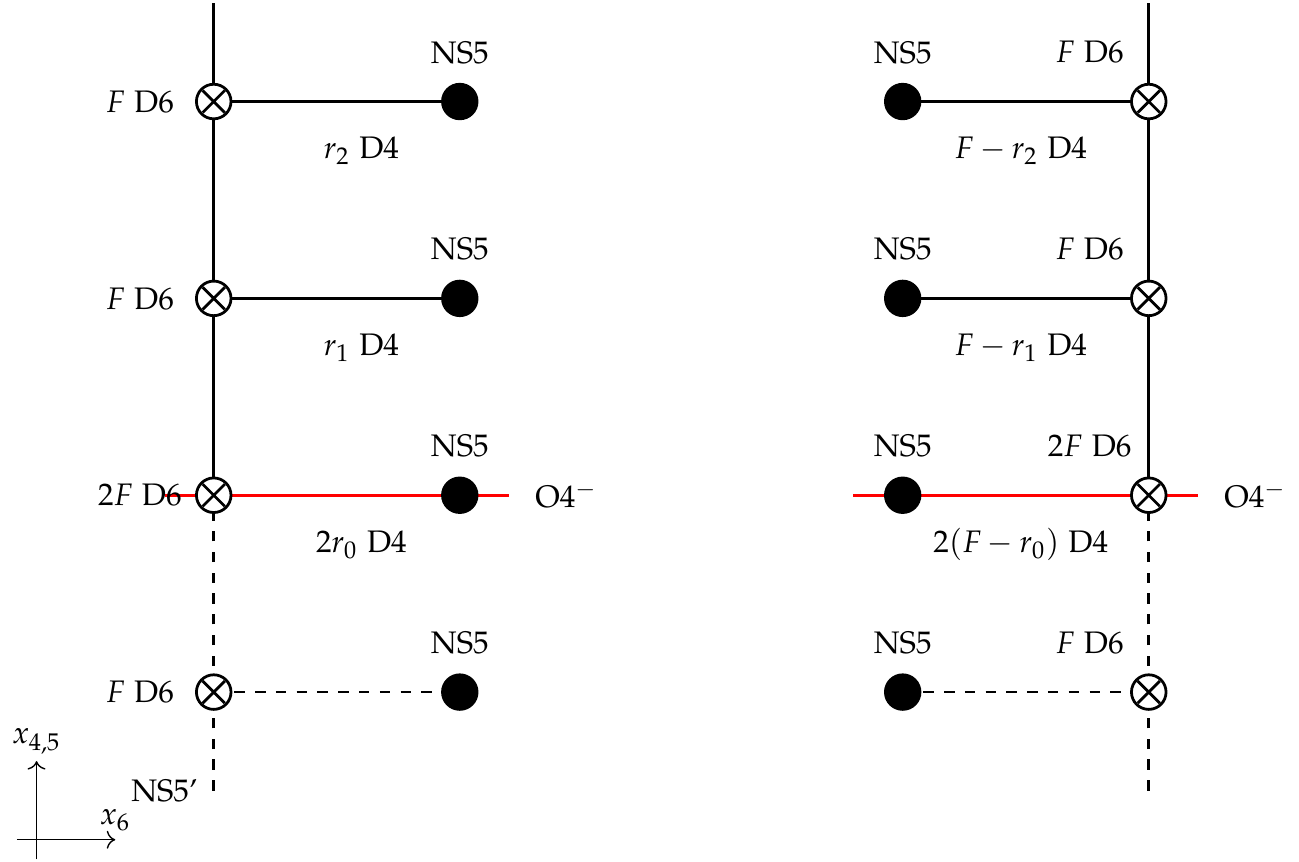}
  \caption{Electric and magnetic sides of the duality for \(Sp(2N)\) gauge theories with adjoint matter.}
  \label{fig:Sp-adjoint}
\end{figure}

In a generic vacuum the adjoint $X$ acquires a vacuum expectation value. At matrix level there is a  rank$ = 2r_0$ sector
at zero vev, and it gives rise to an $SP(2r_0)$ gauge group. The other $k$ rank$=r_i$ sectors, where the
vev of the adjoint is non zero, give raise to a set of $U(r_i)$ sectors.  The ranks are chosen such that $\sum_{i=0}^{k} r_i = N$.
This higgsing corresponds to separating the $\D4$--branes along the directions $4$ and $5$ in the brane picture, as in Figure~\ref{fig:Sp-adjoint}.
Eventually, in a generic vacuum, the gauge group is broken as
\begin{equation}
Sp(2N) \rightarrow Sp(2 r_0) \times \prod_{i=1}^{k} U( r_i) \;.
\end{equation}
in the electric theory and
\begin{equation}
Sp(2\tilde N) \rightarrow Sp(2 (F-r_0-2)) \times \prod_{i=1}^{k} U(F-r_i)\;.
\end{equation}
in the magnetic theory.
At the brane level this dual description is obtained  by first separating the \NS{}--branes, 
then performing the \ac{hw} transition and eventually reconnecting them, see Figure~\ref{fig:Sp-adjoint}.


In the case of $Sp(2N)$ gauge theories with $2F$ fundamentals
and adjoint matter with superpotential $W=\Tr X^{2(k+1)}$
one can perform the reduction in the $Sp(2r_0)$ and in the $U(r_i)$
sectors separately.
In each sector, an superpotential $W_\eta$ is generated.
By reconnecting the branes and using the scale-matching relation
one can identify the bare monopoles of the theory with the product $Sp(2r_0) \times U( r_i)$
with the dressed monopoles of the $Sp(2N)$ theory.
In the dual case the situation is similar.
First one dualizes each sector, obtaining $Sp(2(F-2-r_0))\times \prod U((F-r_i))$, then
reconnects the branes and eventually uses the scale matching relation
to recover the duality without the polynomial deformation in the adjoint field.

We can flow to the Aharony-like duality.
Let us consider $2F+2$ \D5--branes in the electric phase. 
The dual gauge group is broken to
\begin{equation}
Sp(2 \widetilde r_0) \times \prod_{i=1}^{k} U(\widetilde  r_{i }+1)\;,
\end{equation}
where
$\widetilde r_0 = F-r_0-1$ and $\widetilde r_i =  F-r_i$.
In the brane description, we move two \D5--branes in each sector and perform the \ac{hw} transition.
The dual gauge group at $x_3=0$ becomes
\begin{equation}
Sp(2 (F-r_0-1)) \times \prod_{i=1}^{k} U(F - r_i)\; .
\end{equation}
This is given by imposing in the field theory description the correct vacuum structure preserving the duality.
By joining the \NS{}--branes back it becomes
$Sp(2 ((2k+1) F-N-1))$.

As a check we flow to a known duality.
We can consider $2(F+K)$ fundamentals, integrating out $2K$ of them generating a \ac{cs} term. 
One obtains the duality of Kapustin, Kim and Park~\cite{Kapustin:2011vz},
 between an $Sp(2N)_{2K}$-- and an $Sp(2((2k+1)(F+|K|)-N-1)_{-2K}$ gauge theory,
with superpotential as in~\cite{Kapustin:2011vz}.

\section{U(N) groups and antisymmetric matter}
\label{sec:unitary-antisymmetric}
For unitary groups with tensor matter there are two main cases: antisymmetric and symmetric tensors.
We refer the reader to~\cite{Intriligator:1995ax} where these dualities have been first presented.
Here we focus on the antisymmetric case. In the antisymmetric case one has:
\begin{itemize}
\item An $SU(N)$ gauge theory with an antisymmetric tensor $A$, its conjugate $\widetilde A$ with
\begin{align}
\label{asW}
W= \Tr(A \widetilde A )^{2}
\end{align}
and  $F$ flavors is dual to
an $SU(3F-N-4)$ with superpotential
\begin{align}
W=\Tr (a \widetilde a)^{2} + M_1 q \widetilde q + M_0 q \widetilde a a \widetilde q
+ P q  \widetilde a q + \widetilde P \widetilde q a \widetilde q \; ,
\end{align}
where $a,\widetilde a$ are the dual antisymmetric fields,
$q, \widetilde q$ the dual quarks and the mesons are
\begin{align}
P = Q \widetilde A Q \; , && \widetilde P = \widetilde Q A \widetilde Q \; , && M_0 = Q \widetilde Q \; , && M_1 = Q \widetilde A
A \widetilde Q \; .
\end{align}
\item We can also consider the superpotential
\begin{equation}
\label{asW2}
W= \Tr(A \widetilde A )^{2}+ A \widetilde Q \widetilde A Q + (Q \widetilde Q)^2
\end{equation}
in the electric case. The \(SU(N)\) gauge theory
is dual to
an $SU(2 F-N-4)$ gauge theory
with superpotential
\begin{equation}
W=\Tr (a \widetilde a)^{2} + q \widetilde a \widetilde q a +(q \widetilde q)^2 \,.
\end{equation}
This duality can be obtained from the previous one by a Higgs mechanism: a dual meson appears as a linear perturbation in the dual theory. After higgsing
we obtain (\ref{asW2}) from (\ref{asW}) and the dual rank is modified accordingly.\\
\item The discussion can be generalized to the superpotential
$W = \Tr(A \widetilde A)^{k+1}$. In this case one can break the gauge group by
adding a polynomial superpotential in $(A\widetilde A)^{j}$. By turning this superpotential off
one then finds a generalized \ac{kss} duality with dual rank $\widetilde N= (2k+1)F-N-4$
for the generalization of (\ref{asW}) and $\widetilde N= 2k F-N-4$
for the generalization of (\ref{asW2}).
\end{itemize}

\subsection{Brane description}

\begin{figure}
  \centering
  \includegraphics[width=.8\textwidth]{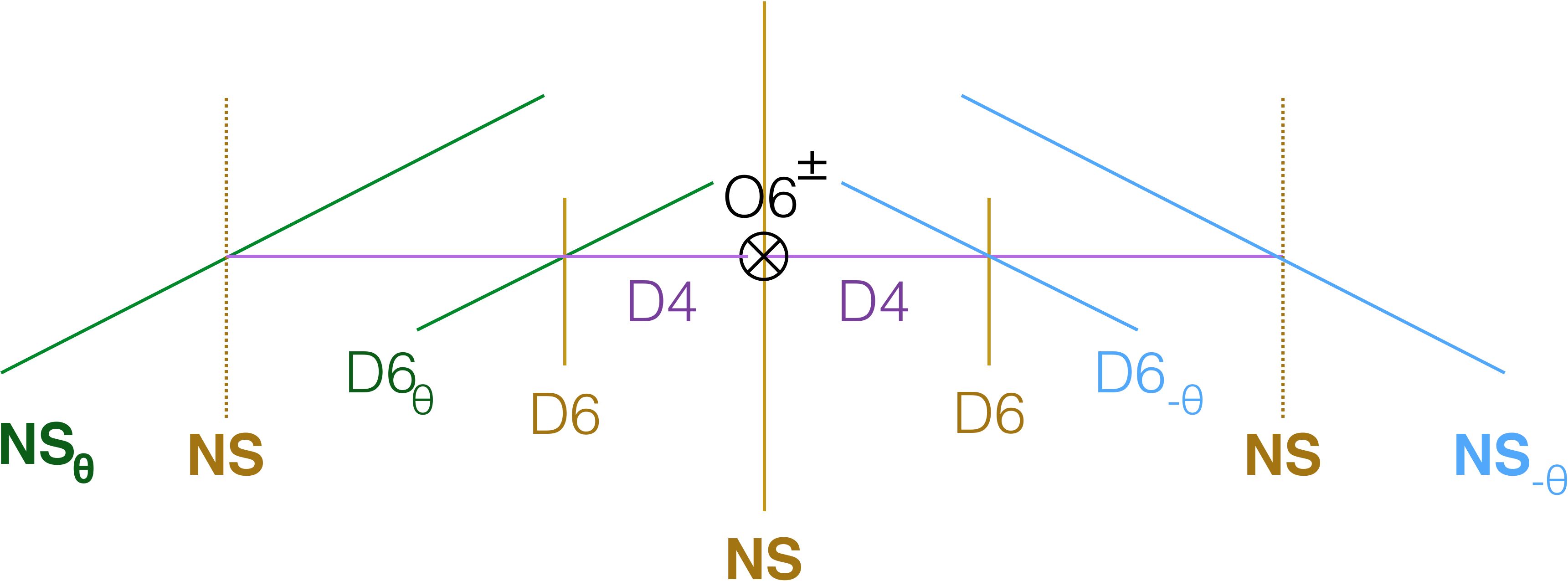}
  \caption{Brane cartoon summarizing all the constructions of unitary
    gauge theories with tensor matter}
  \label{fig:branes-unitary-tensor-matter}
\end{figure}

\begin{figure}
  \centering
  \includegraphics{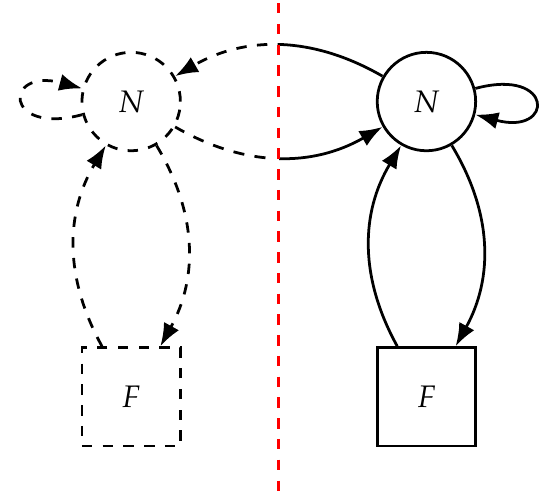}
  \caption{Orbifold projection of the \(A_2\) quiver realizing a U(N) theory with tensor matter.}
  \label{fig:quiver-unitary-tensor-matter}
\end{figure}

The brane realization of these models has been done in~\cite{Landsteiner:1997ei,Csaki:1998mx}.  All cases in this family  can be summarized in the brane cartoon in Figure~\ref{fig:branes-unitary-tensor-matter}.
In order to understand the action of the orientifold we start by discussing a configuration with three \NS--branes without the \O6--plane. The theory is an $\mathcal{N}=2$ quiver with two unitary nodes, connected by a pair of bifundamentals and adjoints (see Figure~\ref{fig:quiver-unitary-tensor-matter}).  At each node there are $F$ flavors.  This configuration and its generalization to $\mathcal{N}=1$ where extensively studied in~\cite{Brodie:1997sz}.
Adding the orientifold plane the two nodes are identified and projected to a single $U(N)$ gauge node. The matter fields are identified as well and there are two possibilities, corresponding to the different signs of the orientifold projection: the pair $(A ,\widetilde A)$ or $(S,\widetilde S)$.  Here we focus on the case with  $(A ,\widetilde A)$.

Now we can break to $\mathcal{N}=1$ by rotating the external \NS{}--branes: rotating the left and right \NS--brane by an angle $\theta$ (\emph{resp.} $-\theta$)
corresponds to introducing a mass term \(\mu(\theta_{\pm} )\) proportional to \(\tan (\theta_{\pm})\) for the adjoints in the $\mathcal{N}=2$ vector multiplet.
Integrating out the massive adjoints
we obtain the superpotential $W = \Tr A \widetilde A$.
If the rotation angle is $\pi/2$, the adjoint is infinitely massive and the superpotential vanishes.
More in general we can consider two stacks of $k$ $\NS{}_{\pm \theta}$--branes, obtaining the superpotential $W =( A\widetilde A)^{k+1}$.

The flavor branes can be added in two ways. In the first case
one can add two stacks of \D6s parallel to the orientifold and to the \NS{}--brane,
one on the left and one on the right.
In the second case one can rotate the stack of \D6s on the left (right) to a stack of
$\D6_{\theta}$ ($\D6_{-\theta}$).
In the first case we have to add the term  $Q \widetilde A \widetilde Q  A + (Q \widetilde Q)^2$ to the superpotential.
In the second case, the flavor branes are parallel to the $\NS_{\pm \theta}$--branes and the quartic terms for the fundamentals are absent.
The two configurations with \D6 or \D6$_{\pm \theta}$--branes have different  ranks in the dual \ac{hw} picture.

\bigskip

The Seiberg duality can be studied in terms of brane motions. It is convenient to describe the motion at first without the orientifold and then add the projection at the end.
The starting theory has two unitary gauge groups connected by a pair of bifundamentals and extra flavors.
The Seiberg-dual phase is obtained by a cascading process, first we dualize one gauge group, then the other, and finally we dualize again the first gauge group.
In terms of branes it corresponds to exchanging the first two \NS{}--branes, then the last two and then the first two again.
Before this exchange, it is convenient to move the \D6--branes.
We have to distinguish the two situations, where we have either two
stacks of $\D6_{\pm \theta}$ 
or two stacks of \D6s parallel to the central brane.
\begin{itemize}
\item
In the first case, the $\D6_\theta$ crosses first the \NS{}--brane and then the $NS{}_{-\theta}$--brane. Both times the crossing generates a stack of \D4--branes.
The same operation has to be performed on the second brane. In this case there are $2$ \D4--branes ending on each $\D6_{\pm \theta}$. The S--rule is not violated because one stack of \D4s is attached to an \NS{}--brane and the other to an $\NS_{\mp \theta}$.
If we interchange the position of the \NS--branes we obtain the dual picture.
The reduction of this duality has been studied in~\cite{Amariti:2015yea} from the brane perspective.

At this point we can consider the effect of the $\O6^-$ orientifold on the central \NS{}--brane.
The following happens.
\begin{enumerate}
\item the gauge group is projected from $SU(N) \times SU(N)$ to $SU(N)$;
\item the bifundamentals connecting the gauge groups become the tensor matter fields;
\item the two flavor groups are identified.
\end{enumerate}
At the level of the duality the orientifold carries the charge of $- 4$ \D6--branes.
By carefully considering the orientifold charge in each transition we end up with the
$SU(3F - N - 4)$ gauge theory as expected.
\item
In the second case the \D6--branes are parallel to the \NS{}--brane. We move the \D6
on the left of the \NS{} towards the $\NS{}_\theta$ and the other in the opposite direction.
Once they cross the $NS_{\pm \theta}$, each \D6 generates a stack of \(F\) \D4--branes.
After this motion the duality works as in the case above. By carefully adding the orientifold charge, the dual $SU(2F-N-4)$ gauge theories are recovered.
\end{itemize}

One can also study the duality with \(k\) $\NS_{\pm \theta}$--branes. In this case one first separates these branes along the direction orthogonal to the plane that they occupy in $(4589)$
and then studies the duality in each sector separately. By reconnecting the branes the
expected dualities are recovered.

\subsection{Dimensional reduction}

Now we compactly $x_3$ and T--dualize along this direction.
We consider the $U(N)$ case, where the baryonic symmetry is gauged.
On the T--dual circle, the theory develops a superpotential of the form
\begin{equation}
\label{anp}
W =  \eta Y_+ Y_-\;,
\end{equation}
where $Y_+=e^{\sigma_1/e_3^2+i \phi_i}$
and
$Y_-=e^{-(\sigma_{N}/e_3^2+i \phi_{N})}$.
This can be understood from the brane picture as follows: there are two sets of \D3s, one connecting
the \NS$_{\theta}$ and \NS{}--branes and the other connecting
the \NS$_{-\theta}$ and \NS{}--branes.
On each stack a superpotential $W_\eta$ is generated by the Euclidean \D1--branes.
The two superpotentials are identical and identified by the orientifold.
Finally, one has (\ref{anp}).

Now we want to investigate the dual phase.
As discussed above there are two possible situations:
the \D5--branes are parallel to the \NS{}--branes \emph{or} to the $NS_{\pm \theta}$--branes.
In the first case $\widetilde N = 3 F-N-4$,
while in the second case we have $\widetilde N = 2 F-N-4$.

On the circle, the extra  superpotential
\begin{equation}
\label{anp2}
W =  \eta' y_+ y_-
\end{equation}
is generated. Here
$y_+=e^{\widetilde \sigma_1/{\widetilde e_3}^2+i \widetilde \phi_i}$
and
$y_-=e^{-(\widetilde \sigma_{\widetilde N}/{\widetilde e_3}^2+i \widetilde \phi_{\widetilde N})}$.
\begin{figure}
  \centering
  \includegraphics[width=.8\textwidth]{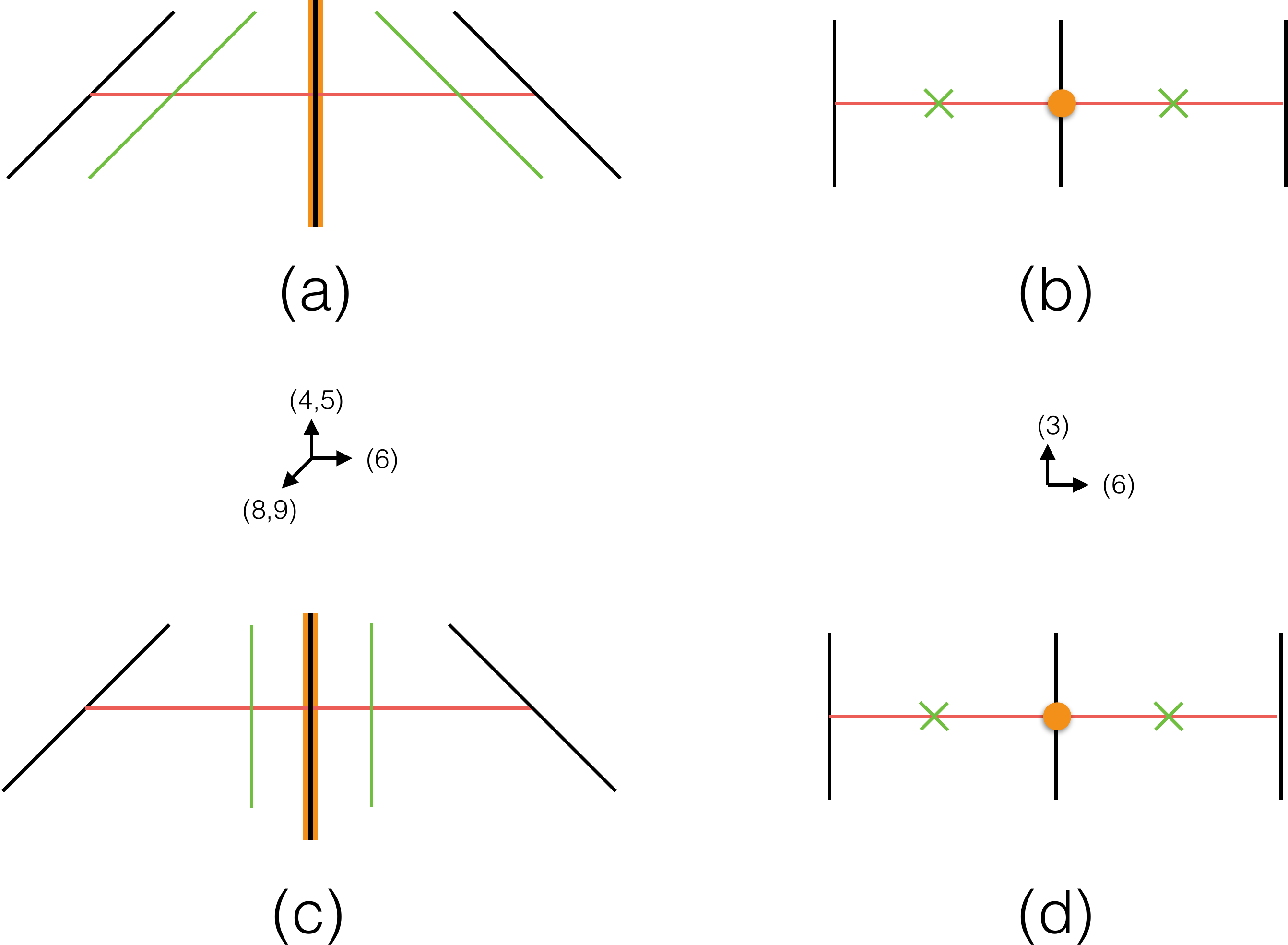}
  \caption{$U(N)$ gauge theory with antisymmetric matter, electric theory. (a) and (b) show the case without superpotential (\D5--branes parallel to $NS_{\pm \theta}$), (c) and (d) show the case with superpotential (\D5--branes are parallel to the \NS{}--branes).}
  \label{fig:lSUAs3d}
\end{figure}
Figures~\ref{fig:lSUAs3d}~(a) and~\ref{fig:lSUAs3d}~(b) show the brane cartoon of the electric theory in the case without the extra superpotential.
The \NS{}--branes are drawn in black, the \D5s in green, the orientifold plane is orange and the \D3s are red.
In Figures~\ref{fig:lSUAs3d}~(c) and~\ref{fig:lSUAs3d}~(d) we represent the case with the superpotential turned on.
Now we want to flow to the theory without the superpotential $W_\eta$.
We consider the case with $F+2$ green branes in each sector, and assign a positive
large real mass to one flavor and one negative large real mass to a second one.  We rotate one pair of $\D5_{\pm \theta}$ clockwise on the circle
and another pair counterclockwise. Finally, we reconnect the pairs at $x_3 = x_3^\circ$, where the second orientifold is placed.

\begin{figure}
  \centering
  \begin{tabular}{cc}
    \includegraphics[width=.7\textwidth]{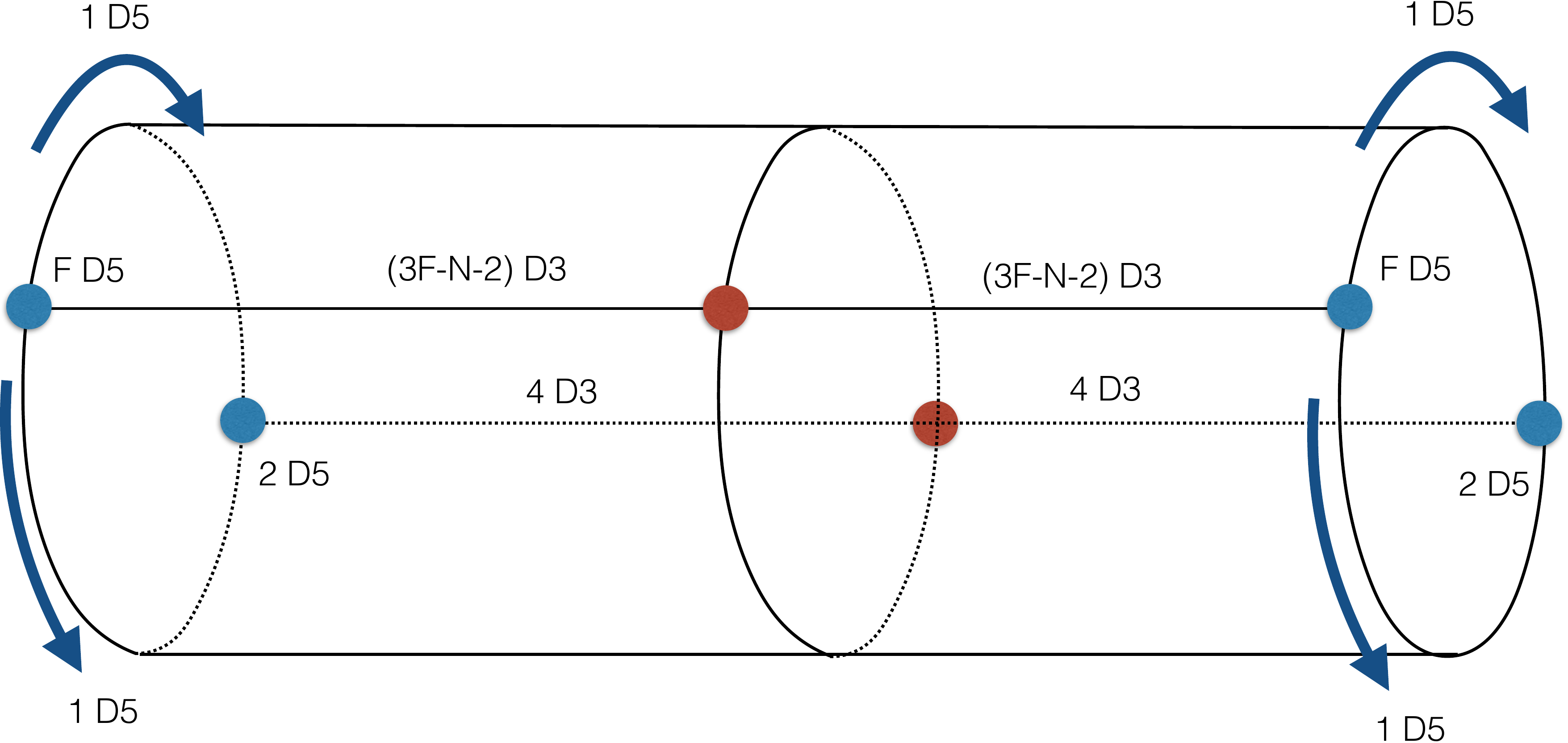}\\
    \includegraphics[width=.7\textwidth]{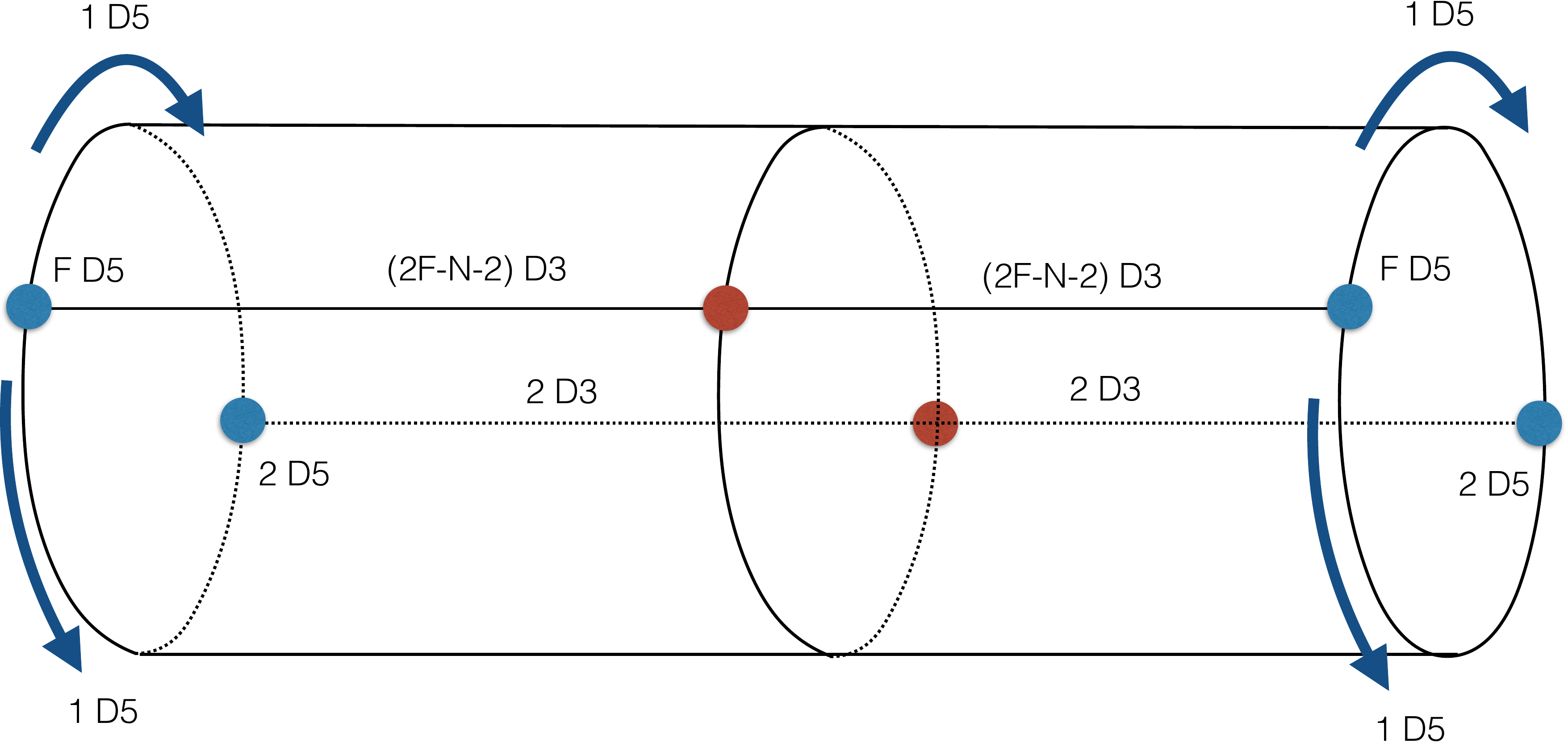}
  \end{tabular}
  \caption{Aharony-like duality for models with antisymmetric matter.}
  \label{fig:32F}
\end{figure}

Now we can proceed as above, we interchange the \NS{}--branes and arrive at the dual configuration.
Finally, we obtain the setup in Figure~\ref{fig:32F}.

These pictures represent the Aharony-like duality for the models with antisymmetric matter.
The extra sectors are dualized to singlets, as done in~\cite{Amariti:2015yea} for the $U(N)$ \textsc{sqcd}.
The extra singlets that are generated interact with the monopoles of the magnetic theory, and they are identified with the Coulomb branch variable of the electric theory. This can be explicitly verified on the field theory side.
The duality now involves a $U(3F-N-2)$ gauge group in the case where the
superpotential is $W = (A \widetilde A)^2$. One can add the extra deformation
$(Q \widetilde Q)^2 + A \widetilde Q \widetilde A Q$ (corresponding to rotating the branes as in Figure~(c)).
In the dual phase this deformation enforces a Higgs flow to the theory with $U(2F-N-2)$,
and it exactly corresponds to the expected dual, after dualizing the extra sectors and considering the \ac{ahw} superpotential.
This confirms the validity of our rules and of our picture.
We can reproduce the same story by considering $k$ external $\NS_\pm{\theta}$--branes.
In this case we can break the $\NS_{\pm{\theta}}$--branes, \emph{e.g.} generating a power superpotential
$W \simeq \sum_i \lambda_i (A \widetilde A)^{j}$.
By breaking the gauge group  in the decoupled sectors we can use the same rules
used above and reconstruct the dual theory.
Finally, we obtain the dual ranks $U((2k+1) F-N-2k)$ and $U(2 k  F - N-2k)$.
As a final check, we can flow to the case with \ac{cs} terms. In this case we reproduce the
duality between the $U(N)_{K}$ theory with $F$ flavors
and the dual $U((2k+1)(F+K)-N-2k)_{-K}$ studied in~\cite{Kapustin:2011vz}.

\section{Orthogonal gauge groups}
\label{sec:orthogonal}
In this section we discuss orthogonal gauge groups.
An  $SO(N)$ gauge theory with $2F$ fundamental vectors and global symmetry $SU(2F) \times U(1)_A \times U(1)_R$ without superpotential
is dual to
an $SO(2F-N-4)$ gauge theory with $2F$ fundamental vectors $q$ and a meson in the (conjugate) symmetric representation of the global $SU(2F)$ with superpotential $W= M q  q $.
This duality was first presented in~\cite{Intriligator:1995id}.
The field content is given in Table~\ref{tab:field-content-SO}.
\begin{table}
  \centering
  \begin{tabular}{ccccccc}
    \toprule
          & \(SO(N)\) & \(SO(\widetilde N)\) & \(SU(2F)\)       & \(U(1)_A\) & \(U(1)_R\)     \\ \midrule
    \(Q\) & \(N\)     & \(1\)                & \(2F\)           & \(1\)      & \(1-(N-2)/F\)  \\
    \(q\) & \(1\)     & \(\widetilde N\)     & \(2\overline F\) & \(-1\)     & \((N-2)/F\)    \\
    \(M\) & \(1\)     & \(1\)                & \(F(2F-1)\)      & \(2\)      & \(2-2(N-2)/F\) \\ \bottomrule
  \end{tabular}
 
  \caption{Field content for the \( SO(N) \) gauge theory with global \(SU(2F) \times U(1)_A \times U(1)_R \) symmetry.}
  \label{tab:field-content-SO}
\end{table}

\subsection{Aspects of field theory}

On can associate three distinct gauge groups to the Lie algebra $so(N)$,
as discussed in~\cite{Aharony:2013hda} where they were called $SO(N)_{\pm}$ and $\Spin(N)$.
In four dimensions the different choices depend on the spectrum of line defects,
while in three dimensions they depend on the
monopole charges in the dual algebra.

The Coulomb branch variables associated to the $so(N)$ algebra are
\begin{equation}
Y_i = e^{(\sigma_i-\sigma_{i-1})/{e_3^2}+ i (\phi_i-\phi_{i-1} )} \; , \quad \quad i=1,\dots, N-1
\end{equation}
and
\begin{equation}
  \begin{cases}
    Y_{N} = e^{(\sigma_{N-1}-\sigma_N)/e_3^2+i(\phi_{N-1}-\phi_N)} & \text{\(N\) even} \;,\\
    Y_{N} = e^{2 \sigma_N/e_3^2+2 i  \phi_N} & \text{\(N\) odd} \;.
  \end{cases}
\end{equation}
 At finite radius there is also a superpotential $W_\eta = \eta Z$
from the \ac{kk} monopoles~\cite{Davies:2000nw,Aharony:2011ci,Aharony:2013kma},
where $Z = Y_1 \prod_{i=2}^{N-2} Y_{i}^2  Y_{N-1} Y_{N}$
in the even case and
$Z = Y_1 \prod_{i=2}^{N-1} Y_{i}^2   Y_{N}$
in the odd case. The two expressions finally boil down to
$Z = e^{(\sigma_1+\sigma_2)/e_3^2+ i (\phi_1+\phi_2)} $.
In presence of matter fields this superpotential still contributes to the theory, but
there is a difference with the symplectic and unitary cases: the superpotential $W_\eta$
does not completely lift the Coulomb branch, parameterized
by $Y_{Spin}= e^{2 \sigma_1/e_3^2+2 i \phi_1}$ in the $\Spin(N)$ case and
 $Y= e^{\sigma_1/e_3^2+i \phi_1}$ in the $SO(N)$ case.
There are three possible dualities:
$\Spin(N) \leftrightarrow SO(\widetilde N)_{-}$, $SO(N)_- \leftrightarrow \Spin(\widetilde N)$ or $SO(N)_+ \leftrightarrow SO(\widetilde N)_{+}$, where in each case $\widetilde N=F-N+2$.

It is possible to reduce the $4$D dualities to $3$D dualities by considering the limit $r \rightarrow 0$, \emph{i.e.} $\eta \rightarrow 0$,
without adding real masses.
This is possible because of the presence of a Coulomb branch.
A region near the origin of the moduli space on the electric side of $SO(N)_+$
corresponds in the dual to the region $\widetilde Y =i/\sqrt{ \widetilde \eta}$.
This breaks the gauge symmetry to
$SO(F-N+2) \times SO(2)$. This last sector in the \textsc{ir} is described by its Coulomb branch variable interacting with the monopole of the unbroken sector through an \ac{ahw} superpotential.
Similarly, one obtains a duality between $SO(N)$ and $\Spin(N)$ theories in pure $3$D.
At the local level, the $O(N)$ duality studied in~\cite{Benini:2011mf,Aharony:2011ci} is recovered.

\subsection{Brane description}
\begin{figure}
  \centering
  \includegraphics{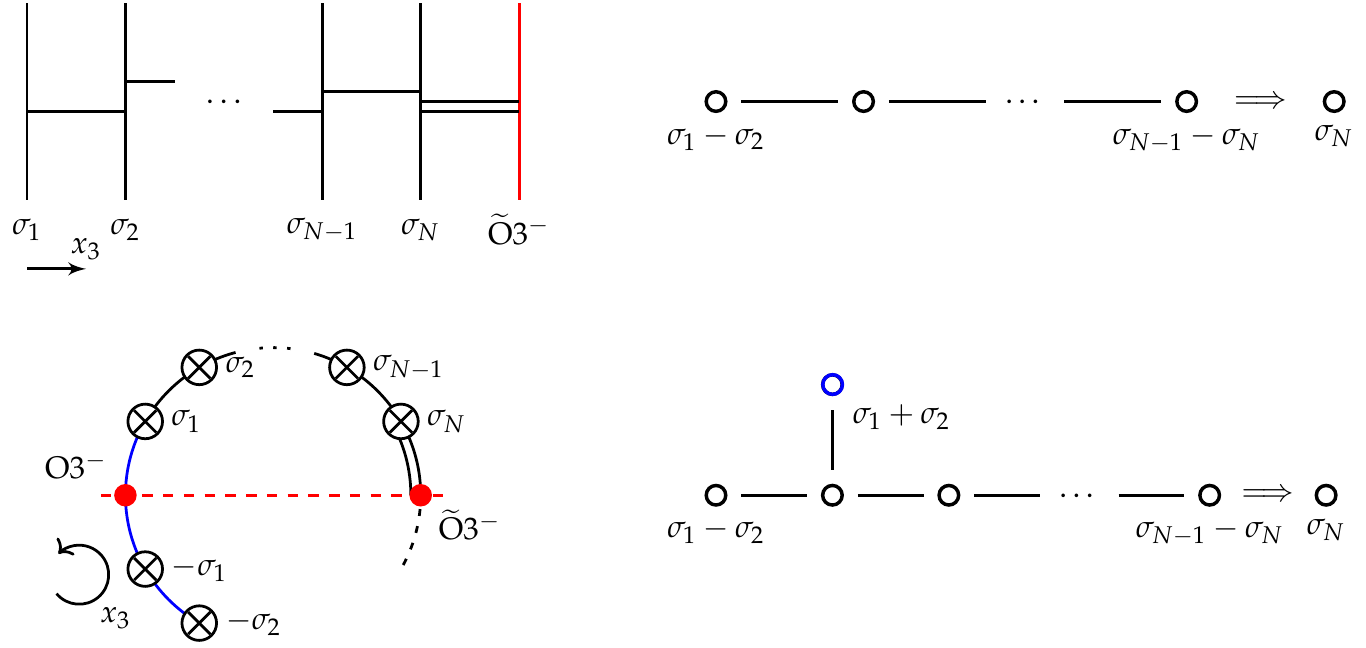}
  \caption{Dynkin and affine Dynkin diagrams and spectrum of \acs{bps} \F1--strings
  associated to the fundamental monopoles for $SO(2N+1)$ theories
  (\(B_N\) algebra). The orientifold at the mirror point \(x_3 = x_3^\circ\) is an \(\O_3^-\), while the one at \(x_3 = 0\) is an \(\widetilde \O_3^-\). For this reason the Dynkin diagram of \(\widetilde B_N\) does not have a \(\setZ_2\) symmetry.}
  \label{fig:Dynkin-Bn}
%
  \centering
  \includegraphics{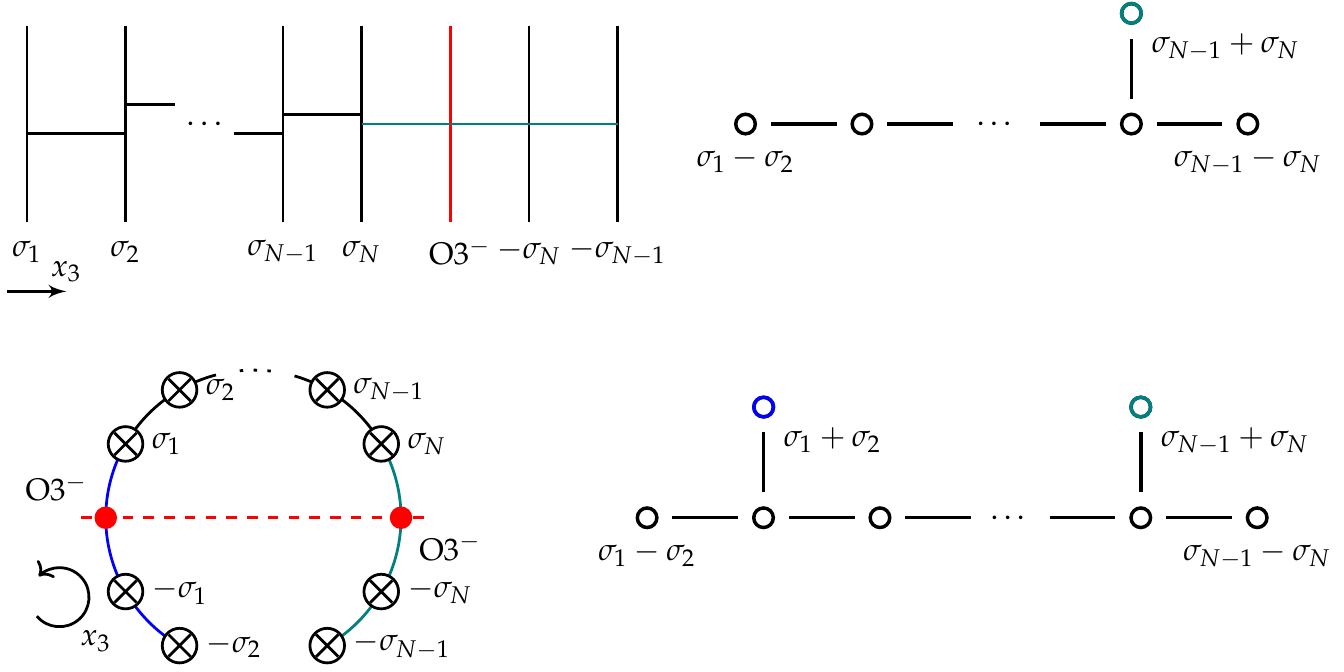}
  \caption{Dynkin and affine Dynkin diagrams and spectrum of \acs{bps} \F1--strings
  associated to the fundamental monopoles for $SO(2N)$ theories
  (\(D_N\) algebra). The affine root is in blue, the root due to the orientifold in \(x_3 = 0\) in teal. Both orientifolds are \(\O_3^-\) and the affine Dynkin diagram has a \(\setZ_2\) symmetry.}
  \label{fig:Dynkin-Dn}
\end{figure}
Now we turn to the brane picture. We will limit the discussion to the study of the local properties, without focusing on the difference between the $\mathrm{(S)pin}(N)$ and the  $(S)O_\pm$ cases.
We will comment on the possibility of extending the analysis to the global property of the gauge group in the conclusions.
At the brane level these theories are obtained in two different ways. In one case, we put the $O4^{-}$ ($\widetilde O4^{-}$) on $N=2n$ ($N=2n+1$) \D4--branes stretched between the \NS{} and \NS'--brane. This theory has an $Sp(2F)$ global symmetry and
we expect that this symmetry is enhanced to $SU(2F)$.
 In the second case we consider two $\NS{}_{\pm\theta}$--branes connected by a stack of \D4--branes intersecting the $\O6^+$--plane symmetrically with respect to the $\NS{}_{\pm \theta}$--branes.
We can distinguish between the even $N=2n$ case and the odd $N=2n+1$
case, essentially this corresponds to the number of \D4--branes. In this case the global symmetry is $SU(F)$
and we expect this enhances to $SU(F)^2$.

In the T--dual \tIIB description there is an \O3--plane between the two \NS$_{\pm \theta}$ and the
\(2 F\) $\D5_{\pm \theta}$--branes. The gauge theory lives on the $N$ \D3s extended along $x_6$.
First we study the generation of the superpotential in the Coulomb branch in the case of a pure gauge theory. Then we discuss the new duality obtained on the circle and finally we reproduce the $3$D limit studied in~\cite{Aharony:2013kma}.
The three-dimensional theory on the circle has two possible orientifolds $\O3^-$
or $\widetilde \O3^-$. In the first case we have to consider an even number of \D3--branes while in the second case they have to be odd. We can study in both cases the generation of the superpotential in terms of the Coulomb branch variables.

The superpotential on the Coulomb branch  is obtained in terms of the spectrum of
the allowed \acs{bps} \F1--strings in presence of the orientifold, as discussed in Section~\ref{sec:general}.
In the orthogonal case we can represent the two different possibilities
for the $\O3^-$ or $\widetilde \O3^-$ with the $B_N$ and the $D_N$ series.
(see Figure~\ref{fig:Dynkin-Bn} and Figure~\ref{fig:Dynkin-Dn}).

The extra superpotential corresponds in both cases to the extra term $Z = e^{(\sigma_1+\sigma_2)/e_3^2+ i (\phi_1+\phi_2)}$. Since it identifies two eigenvalues after we cross
from one half to the other of the circle it involves the identification and add the superpotential $W_\eta$.
Finally, we have
\begin{align}
W_{SO(2n)} = \sum_{i=1}^{r_G} \frac{1}{Y_i} + \eta Z \; , &&
W_{SO(2n+1)} = \sum_{i=1}^{r_G-1} \frac{1}{Y_i} +\frac{2}{Y_{r_G}} + \eta Z \;.
\end{align}
When we consider the \D6--branes there is an unlifted direction in the Coulomb branch, corresponding to the term
$e^{2 \sigma_1/e_3^2}$.

We consider $F+2$ \D5 on each \NS{}--brane and take the pure $3$D limit.
In the electric theory we are left with a pure $3$D  $so(N)$ theory with $2F$ flavors.
In the dual theory the situation is more intricate.
At $x_3=0$ there is an $so(F-N-2)$ theory with superpotential $W= M q q $.
At  $x_3=x_3^\circ$ there is an $so(4)$ gauge theory with two fundamentals.
It can be dualized to a singlet Y, 
interacting with the $so(F-N-2)$ through an \ac{ahw} superpotential.
This interaction is $W = y Y$ where  $y$ is the magnetic monopole.  
By interpreting $Y$ as the electric monopole acting as a singlet in the magnetic theory 
we arrive to the expected duality.

\subsubsection{An alternative limit}

Differently from the unitary and symplectic cases, here the pure $3$D limit can be obtained without 
any real mass flow~\cite{Aharony:2013kma}.
The reason is that  the region $x_3\simeq 0$ of the Coulomb branch in the electric theory 
corresponds to the region $x_3 \simeq x_3^\circ$
in the magnetic one. An $so(2)$ gauge theory is created at $x_3^\circ$ in the magnetic theory, and
the pure $3$D limit can be taken directly, preserving the duality.

\begin{figure}
  \centering
 \includegraphics[width=.7\textwidth]{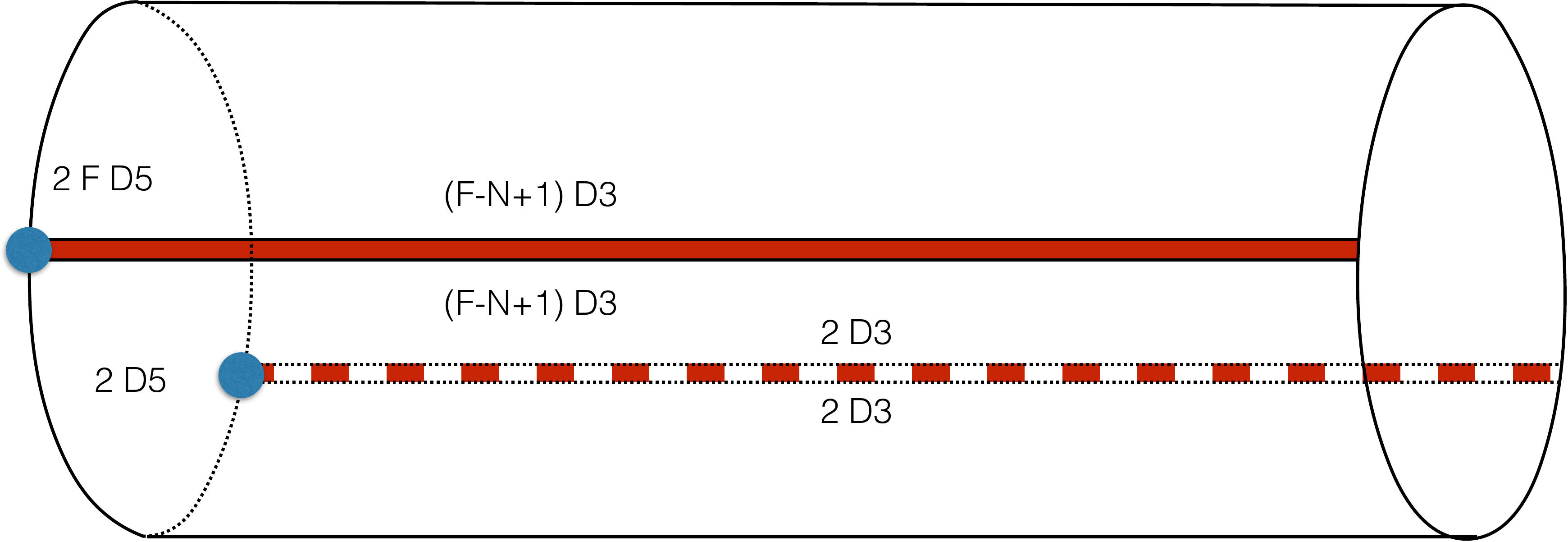}
  \caption{Two \D5--branes reconnect at the mirror point of the circle.}
  \label{fig:scalingO(n)}
\end{figure}

In the brane description we consider the \D3s in the electric theory at the origin, while in the magnetic theory the  orientifold
generates automatically a pair of \D3--branes at $x_3^\circ$.
The dual gauge theory becomes $so(2F-N-2) \times so(2)$. 
The final configuration is in Figure~\ref{fig:scalingO(n)}. 
The $so(2)$ sector
is dual to a singlet $Y$, that interacts with the  $so(2F-N-2)$ sector  through an \ac{ahw} superpotential.
Again, in the pure $3$D case, $Y$ has the same quantum numbers as the electric monopole.

\bigskip

It is possible to study the case with \O6--planes as well. In this case the discussion
follows the one of the symplectic case, and we do not report the whole analysis.
The mirror orientifold is created on the circle and the extra sectors can be studied with the
usual brane techniques. One can also study the cases with tensor matter, by adding
$k$ \NS'--branes in the case with an \O4--plane and $k$ NS$_{\pm \theta}$ for the cases with
the \O6--planes. Moreover, one can consider the cases with \O6--planes and
an extra \NS{}--brane, this leads to unitary theories with symmetric matter and
the discussion follows the one in Section~\ref{sec:unitary-antisymmetric}.
In all the cases new examples of three-dimensional dualities can be worked out.
We conclude by observing that the known three-dimensional case studied in~\cite{Kim:2013cma} can be recovered
from these dualities.

\section{Conclusions}
\label{ref:conclusions}

In this paper we completed the analysis started in~\cite{Amariti:2015yea} of the reduction
of four-dimensional dualities to three dimensions via brane constructions.
We have shown that this picture captures the relevant properties
of the reduction of the duality on $\mathbb{R}^3 \times S^1$.
By T--duality the Coulomb branch on the circle is correctly described, after separating the
\D3--branes in the compact direction, by an affine Toda potential for the \F1--strings in an S-dual frame.
When considering real groups or tensor matter fields,
a crucial role is played by the behavior of the orientifold
under T--duality. A second orientifold plane is generated
at an opposite point on the T--dual circle.
We have shown that it is necessary to consider the physics at this mirror
point when taking the three-dimensional limit.
This limit is a double scaling on the real masses and the radius.
The masses correspond to the positions of certain \D5--branes (and in the magnetic phases
also \D3--branes). By reconnecting the branes at the mirror point, a new unified scenario
to study the reduction of four-dimensional dualities admitting a \tIIA description in four dimensions emerges.
The construction presents an algorithmic way to obtain many new three-dimensional dual pairs
from their four-dimensional parents which we have discussed in this article.

\bigskip

The construction presented here is generic for $4$D dualities that can
be described by \tIIA brane systems and several extensions are possible.
\emph{E.g.\ }one could apply the reduction to \tIIA setups involving chiral matter and orientifolds
like the ones studied in~\cite{Hanany:1997sa,Hanany:1997tb,Hanany:1999sj}.

It would be interesting to study the spectrum of line defects and their connection to dualities in the brane picture.
The relations of line defects to global properties of the gauge groups has been pointed out~\cite{Aharony:2013hda}
and there are various implications for the duality involving the orthogonal algebras $so(n)$~\cite{Aharony:2013kma}.
It should be possible to distinguish between ``$\Spin(N)$'' and the ``$SO(N)_\pm$'' (in the language of~\cite{Aharony:2013kma})
also in the brane setup, \emph{e.g.\ }following the discussion in~\cite{Moore:2014gua}.
More explicitly, in the \tIIB description,
one can separate the \D3--branes, studying configurations of
semi-infinite  (electric) \D1--branes and (magnetic) \F1--strings with endpoints on the \D3--branes.
In the presence of an orientifold, this analysis should give rise to the distinction between $\Spin(N)$ and $SO(N)_\pm$.
We leave this problem for future investigation.

Another interesting extension of our analysis involves the pairs of
orientifolds associated to \emph{twisted} affine Dynkin diagrams.
These cases do not descend from a compactification of a \tIIA background, and they do not represent a four-dimensional theory.
However, they do correspond to well-defined theories on $\mathbb{R}^3 \times S^1$.
One might expect obtaining new Seiberg-like dualities corresponding to these configurations.
By assigning suitable real masses one may even expect to obtain new purely three-dimensional dualities, without four-dimensional parents.
It would be interesting to further investigate in this direction.

Let us comment on the generation of the monopole charges.
The axial $U(1)_A$,
anomalous in four dimensions, is broken by the \ac{kk} monopole superpotential at finite radius.
However, rotating the \D5s on the circle partially breaks the non-abelian flavor symmetry and generates the axial symmetry.
The massless singlets located at $x_3 = x_3^\circ$ are charged under this symmetry and survive the pure $3$D limit. 
They correspond to the monopoles of the electric theory and, at the same time,
their $U(1)_A$ charge is imposed by the original global symmetry.
This observation explains  the relation pointed out in~\cite{Benna:2009xd}, 
between the equations governing the cancellation of the anomalies 
in $4$D and those governing the monopole charges in $3$D.

It is possible to reproduce our results when reducing
the four-dimensional superconformal index~\cite{Romelsberger:2005eg,Kinney:2005ej}
to the
three-dimensional partition function~\cite{Jafferis:2010un,Hama:2010av}.
That was done for the case of \textsc{sqcd} in~\cite{Aharony:2013dha} and in presence of adjoint matter in~\cite{Amariti:2014iza} \footnote{
Another example, involving matter matter in antisymmetric representation, appeared in~\cite{Gahramanov:2013pca}.}.
One should consider the identities summarized in~\cite{Spiridonov:2009za} and obtain new identities
for the three dimensional dualities.
A possible strategy for this calculation is the \ac{kk} reduction of the one-loop determinants
while shifting some fugacities of the global and local symmetries.
This reproduces the double-scaling limit discussed in this paper.
One should check that the surviving zero modes remove the possible divergent contributions found in~\cite{Aharony:2013kma}.

\section*{Acknowledgments}
\begin{small}
  The authors would like to thank Alberto Zaffaroni, Francisco Morales, Jan Troost, and Massimo Bianchi  for
   discussions and comments.

   \noindent
   A.A. is funded by the European Research Council
    (\textsc{erc}-2012-\textsc{adg}\_20120216) and
    acknowledges support by \textsc{anr} grant
    13-\textsc{bs}05-0001. D.F. is
    \textsc{frs-fnrs} Charg\'e de Recherches. He acknowledges support
    by the \textsc{frs-fnrs}, by \textsc{iisn} - Belgium through
    conventions 4.4511.06 and 4.4514.08, by the Communaut\'e Francaise
    de Belgique through the \textsc{arc} program and by the
    \textsc{erc} through the SyDuGraM Advanced Grant.  C.K.
    acknowledges support by \textsc{anr} grant 12-\textsc{bs}05-003-01
    and by Enhanced Eurotalents, which is co-funded by \textsc{cea}
    and the European Commission. The work of S.R. is supported by the
    Swiss National Science Foundation (\textsc{snf}) under grant
    number \textsc{pp}00\textsc{p}2\_157571/1.

\noindent
    D.O. and S.R. would like to thank the Kavli \textsc{ipmu} for
    hospitality during the final stages of this work.
\end{small}

\appendix

\section{Conventions}
\label{App:geometry}
In this appendix we summarize the conventions of the geometry that we used in the paper.
We consider a circle of length $\beta$ and radius $r = \beta/(2 \pi)$.
The kinetic term is normalized as in~\cite{Davies:1999uw}
\begin{equation}
  S = \frac{\beta}{4 g_4^2} F_{\mu \nu} F^{\mu\nu} = \frac{1}{4 g_3^2} F_{\mu \nu} F^{\mu\nu} \; ,  
\end{equation}
where we used the relation  $g_4^2 = 2 \pi r g_3^2$.
The Coulomb branch variables are
\begin{equation}
  X_i = e^{4 \pi \sigma_i/g_3^2 + i \phi_i} \; ,
\end{equation}
with periodicity of $\sigma_i$ proportional to $1/r$.
It follows that
\begin{equation}
\eta\equiv \Lambda^b =e^{-4 \pi /(r g_3^2)}\; .
\end{equation}
To simplify the notation in the paper we work with the coupling $e_3^2 =  g_3^2/(4 \pi)$.

\section{Orientifolds}
\label{sec:orientifolds}

Orientifold planes played a special role in our discussion, therefore we briefly review here some of their basic aspects~\cite{Sagnotti:1987tw,Horava:1989vt,Polchinski:1996na}.
A \(p\)-dimensional orientifold (\O{p}--plane)
is defined in string theory by its perturbative action.
It corresponds to the projection
$ \sigma  \cdot \Omega \cdot (-1)^{F_L}$,
where $\sigma$ is the parity inversion of the coordinates transverse to the plane,
$\Omega$ is the world-sheet
parity and $F_L$ is the left-moving fermion number.
The orientifold acts both on the NS~\cite{Gimon:1996rq} and on the R sector~\cite{Witten:1998xy,Uranga:1998uj}, by two distinct \(\setZ_2\) parities.
The action on the NS sector is perturbative in string theory  and we denote it with a $+$ or a $-$.
The $\setZ_2$ acting on the R sector is non-perturbative in string theory and we denote it
with the presence or the absence of a tilde (\(\sim\)) on the orientifold~\cite{Hanany:1999sj,Hanany:2000fq}.
These charges identify the action of the orientifold on the gauge symmetry.
We summarize the different possibilities in Table~\ref{tab:orientifold-charges}. 
\begin{table}[t]
  \centering
  \begin{tabular}{cccccc}
    \toprule
    type                   & charge        & gauge       \\ \midrule
    \(\O{p}^-\)            & \(-2^{p-5}\)  & \(SO(2n)\)   \\
    \(\O{p}^+\)            & \(2^{p-5}\)   & \(Sp(2n)\)   \\
    \(\widetilde \O{p}^-\) & \(2^{p-5}/2\) & \(SO(2n+1)\) \\
    \(\widetilde \O{p}^+\) & \(2^{p-5}\)   & \(Sp(2n)\)   \\ \bottomrule
  \end{tabular}
  \caption{Orientifold charges and corresponding gauge symmetry}
  \label{tab:orientifold-charges}
\end{table}

In this paper we have been mostly interested in the $\O3$ and $\O5$ cases,
coming from a T--duality from \tIIA.
After compactification an $\O{(p+1)}$ plane becomes a
pair of orientifolds which turn, after T--duality, into a pair of  $\O{p}$--planes (see Figure~\ref{fig:new-orientifold})~\cite{Hanany:1999sj,Hanany:2000fq},
and this fact has been crucial in our analysis.
There are in principle 16 different possibilities, depending on the discrete torsion,
but only some of them have been relevant for our analysis, as we discussed in the paper.

\setstretch{.95}

\printbibliography

\end{document}